\documentclass{article} 
\usepackage{amsmath,amssymb,stmaryrd,amsthm} 
\usepackage{xspace}
\usepackage[latin1]{inputenc}
\usepackage{a4wide}
\usepackage[french,english]{babel}
\usepackage[ps,dvips,arrow,matrix]{xy}%curve,frame,tips
\usepackage{pstricks,pst-node}\newgray{myGray}{0.5}
\usepackage{ulem}

\usepackage{bussproofs} %fourni
\newcommand{\Par}{\bindnasrepma}

\EnableBpAbbreviations

\newcommand{\ssi}{\text{ \iflanguage{french}{ssi}{iff} }}

\newcommand{\et}{\text{ \iflanguage{french}{et}{and} }}
\newcommand{\ou}{\text{ \iflanguage{french}{ou}{or} }}
\newcommand{\ie}{\makebox{\emph{i.e. }}}
\newcommand{\eg}{\makebox{\emph{e.g. }}}

\newcommand{\pcf}{\textsc{pcf}\xspace}
\newcommand{\Rel}{\ensuremath{\mathbf{Rel}}\xspace}
\newcommand{\Bip}{\ensuremath{\mathbf{Bip}}\xspace}

\newcommand{\Pcoh}{\iflanguage{french}{$P$-cohérence\xspace}{$P$-co\-her\-ence\xspace}}

\newcommand{\Kcoh}[1][K]{\iflanguage{french}{$#1$-cohérent\xspace}{$#1$-co\-her\-ence\xspace}}

\newcommand{\uKcoh}[1][K]{\iflanguage{french}{$#1$-cohérent uniforme\xspace}{uniform $#1$-coherence\xspace}}

\newcommand{\bindex}[2][]{\emph{#2}%
\begingroup%
\setbox0=\hbox{#1}%
\setbox1=\hbox{\hskip0.5em}%
\ifnum\wd1<\wd0\index{#1}%
\else\index{#2}\fi%
\endgroup%
}

\newcommand{\bbbn}{\mathbb{N}}

\newcommand{\subfinast}{\subseteq_{\mathrm{fin}}^{\ast}}

\newcommand{\Pfin}{{\mathcal P}_{\mathrm{fin}}}

\newcommand{\Pfinast}{{\mathcal P}_{\mathrm{fin}}^{\ast}}

\newcommand\MultK[1][K]{\mathcal M_{#1}}
\newcommand{\Mfin}{\mathcal{M}_{\text{fin}}}

\newcommand{\supp}[2][]{% A REGLER C'EST MERDIQUE !!!
  \begingroup%
    \setbox0=\hbox{$#2$}
    \setbox1=\hbox{\hskip0.01em}
    \ifnum\wd1<\wd0 \operatorname{supp}^{#1}(#2)
    \else \operatorname{supp}^{#1}\fi
  \endgroup}%

\newcommand{\card}{\sharp}

\newcommand{\UL}[1]{\underline{#1}}
\newcommand{\mathsout}[1]{\text{\sout{$#1$}}}% ``strike out'' bricolé.

\newcommand{\bigK}{\bbbn\backslash\{0,1\}}
\newcommand{\Cl}{\ensuremath{\operatorname{Cl}}}

\newcommand{\Rcoh}[1][X]{\makebox{\setlength{\unitlength}{0.1em}%
\begin{picture}(10,0)
      \put(0,-1.5){$\smile$}\put(0,2){$\frown$}\end{picture}}_{#1}}
\newcommand{\Rcohs}[1][X]{\mathord{\frown}_{#1}}
\newcommand{\Rincoh}[1][X]{\makebox{\setlength{\unitlength}{0.1em}%
\begin{picture}(10,0)
      \put(0,-1.5){$\frown$}\put(0,2){$\smile$}\end{picture}}_{#1}}
\newcommand{\Rincohs}[1][X]{\mathord{\smile}_{#1}}
\newcommand{\Rneutre}[1][X]{\mathsf{N}_{#1}}

\newcommand{\sect}{\triangleleft}

\newcommand{\msect}{\leftslice}%\triangleleft

\newcommand{\hcsem}[1]{%
  \begingroup%
    \setbox0=\hbox{$#1$}
    \setbox1=\hbox{\hskip1em}
    \ifnum\wd1<\wd0 (#1)^{\bullet}
    \else #1^{\bullet}\fi
  \endgroup
}%autre possibilité pompeuse : ^{\bullet\text{hc}}

\newcommand{\U}[1][]{%U_{K,\emptyset}
\begingroup%
    \setbox0=\hbox{$#1$}
    \setbox1=\hbox{\hskip0.2em}
    \ifnum\wd1<\wd0 U(#1)
    \else U\fi
\endgroup}
\newcommand{\UK}[1][]{%U_{K,\emptyset}
\begingroup%
    \setbox0=\hbox{$#1$}
    \setbox1=\hbox{\hskip0.2em}
    \ifnum\wd1<\wd0 U_{K,\emptyset}(#1)
    \else U_{K,\emptyset}\fi
\endgroup}
\newcommand{\fnu}[1][]{% variante en \nu
\begingroup%
    \setbox0=\hbox{$#1$}
    \setbox1=\hbox{\hskip0.2em}
    \ifnum\wd1<\wd0 \operatorname{nu}(#1)
    \else \operatorname{nu}\fi
  \endgroup}
\newcommand{\fnuc}[1][]{% variante en \nu_2
\begingroup%
    \setbox0=\hbox{$#1$}
    \setbox1=\hbox{\hskip0.2em}
    \ifnum\wd1<\wd0 \operatorname{nu}_2(#1)
    \else \operatorname{nu}_2\fi
  \endgroup}
\newcommand{\fu}[1][]{% variante en \upsilon
\begingroup%
    \setbox0=\hbox{$#1$}
    \setbox1=\hbox{\hskip0.2em}
    \ifnum\wd1<\wd0 \operatorname{u}(#1)
    \else \operatorname{u}\fi
  \endgroup}

\newcommand{\NK}[2][K]{|#2|_{\mathsf{N},#1}}%la trame neutre
\newcommand{\N}[1]{|#1|_{\mathsf{N}}}

\newcommand{\fNK}[1][K]{N_{#1}}
\newcommand{\fN}{N}

\newcommand{\IK}[2][K]{\ensuremath{[#2]_{#1}}}
\newcommand{\IKu}[2][K]{\ensuremath{[#2]^u_{#1}}}

\newcommand{\tramePos}[1]{\ensuremath{|#1|^{+}}}
\newcommand{\trameNeg}[1]{\ensuremath{|#1|^{-}}}
\newcommand{\tP}[1]{\tramePos{#1}}
\newcommand{\tN}[1]{\trameNeg{#1}}
\newcommand{\Pos}{\operatorname{Pos}}

\newcommand{\loli}{\multimap}
\newcommand{\orth}{{^\perp}}
\newcommand{\with}{\mathrel{\&}}

\newcommand{\ue}{\mathop{\underset{u}{!}}}
\newcommand{\uwn}{\mathop{\underset{u}{?}}}

\newcommand{\se}{\mathop{\underset{\iflanguage{french}{e}{s}}{!}}}

\newcommand{\nuhce}{\mathop{\underset{nuh}{!}}}

\newcommand{\tabDP}{\shortstack{\smallskip\\\DisplayProof\smallskip}}
\newcommand{\AXR}[2][\AxRuleName]{%
\AXC{}\RightLabel{#1}\UIC{$\vdash$\ensuremath{#2}}%
}

\newcommand{\AXL}[2][\AxRuleName]{%
\AXC{}\LeftLabel{#1}\UIC{$\vdash$\ensuremath{#2}}%
}

\newcommand{\CUTR}[2][\CutRuleName]{%
\RightLabel{#1}\BIC{$\vdash$\ensuremath{#2}}%
}

\newcommand{\TOPR}[2][\TopRuleName]{%
\AXC{}\RightLabel{#1}\UIC{$\vdash$\ensuremath{#2}}
}

\newcommand{\AVECR}[2][\WithRuleName]{%
\RL{#1}
\BIC{$\vdash$\ensuremath{#2}}
}

\newcommand{\AVECL}[2][\WithRuleName]{%
\LL{#1}
\BIC{$\vdash$\ensuremath{#2}}
}

\newcommand{\PLUSR}[2][\PlusRuleName]{%
\RL{#1}\UIC{$\vdash$\ensuremath{#2}}
}

\newcommand{\BOTR}[2][\BotRuleName]{%
\RL{#1}\UIC{$\vdash$\ensuremath{#2}}
}

\newcommand{\BOTL}[2][\BotRuleName]{%
\LL{#1}\UIC{$\vdash$\ensuremath{#2}}
}

\newcommand{\UNR}[1][\OneRuleName]{
\AXC{}
\RL{#1}
\UIC{$\vdash 1$}
}

\newcommand{\altUNR}[2][\OneRuleName]{
\AXC{}
\RL{#1}
\UIC{$\vdash$\ensuremath{#2}}
}

\newcommand{\PARR}[2][\ParRuleName]{%
\RL{#1}\UIC{$\vdash$\ensuremath{#2}}
}

\newcommand{\TENSR}[2][\TensRuleName]{
\RL{#1}\BIC{$\vdash$\ensuremath{#2}}
}

\newcommand{\DERR}[2][\DerRuleName]{
\RL{#1}\UIC{$\vdash$\ensuremath{#2}}
}

\newcommand{\DERL}[2][\DerRuleName]{
\LL{#1}\UIC{$\vdash$\ensuremath{#2}}
}

\newcommand{\CONTR}[2][\ContRuleName]{
\RL{#1}\UIC{$\vdash$\ensuremath{#2}}
}

\newcommand{\CONTL}[2][\ContRuleName]{
\LL{#1}\UIC{$\vdash$\ensuremath{#2}}
}

\newcommand{\AFFR}[2][\WeakRuleName]{
\RL{#1}
\UIC{$\vdash$\ensuremath{#2}}
}

\newcommand{\PROMR}[2][\PromRuleName]{
\RL{#1}\UIC{$\vdash$\ensuremath{#2}}
}

\newcommand{\DIVERGER}[2][\DivergeRuleName]{
\AXC{}\RL{#1}\UIC{$\vdash$\ensuremath{#2}}
}

\newcommand{\GIVEUPR}[1][\GiveUpRuleName]{
\AXC{}
\RL{#1}
\UIC{$\vdash$}
}

\newcommand{\HC}{\ensuremath{\mathbf{Hc}}\xspace}
\newcommand{\NHC}{\ensuremath{\mathbf{NHc}}\xspace}

\newcommand{\NCOHK}[1][K]{\ensuremath{\mathbf{NCOH}_{#1}}\xspace}
\newcommand{\NCohK}[1][K]{\ensuremath{\mathbf{NCoh}_{#1}}\xspace}
\newcommand{\CohK}[1][K]{\ensuremath{\mathbf{Coh}_{#1}}\xspace}

\newcommand{\I}[1]{%
  \begingroup%
    \setbox0=\hbox{$#1$}
    \setbox1=\hbox{\hskip1em}
    \ifnum\wd1<\wd0 (#1){^{\bullet}}
    \else #1{^{\bullet}}\fi
  \endgroup}

\newcommand{\comp}{\mathrel{\fatsemi}}

\newcommand{\id}[1][]{\operatorname{id}_{#1}}
\newcommand{\Id}{\operatorname{id}}
\newcommand{\der}[1][]{\operatorname{der}_{#1}}
\newcommand{\dig}[1][]{\operatorname{dig}_{#1}}

\newcommand{\C}{\ensuremath{{\cal C}}}

\newcommand{\unit}[1][X]{\operatorname{unit}_{#1}}
\newcommand{\com}[1][X]{\operatorname{com}_{#1}}
\newcommand{\sym}[1][X]{\operatorname{sym}_{#1}}
\newcommand{\ass}[1][X,Y,Z]{\operatorname{ass}_{#1}} %virer les X

\newcommand{\cont}[1][]{\operatorname{cont}_{#1}}

\newcommand{\aff}[1][]{%
\operatorname{\iflanguage{english}{weak}{aff}}_{#1}}

\newcommand{\bool}{\ensuremath{\textbf{bool}}}
\newcommand{\nat}{\ensuremath{\textbf{nat}}}
\newcommand{\true}{\ensuremath{\texttt{\iflanguage{french}vt}}}
\newcommand{\false}{\ensuremath{\texttt{f}}}

\newcommand{\Ifthenelse}[3]{\mathop{\texttt{if}} #1
\mathrel{\texttt{then}} #2 \mathrel{\texttt{else}} #3}

\newcounter{theocount}[section]
\newcounter{definitioncount}[section]

\newcounter{propcount}[section]
\newcounter{corollairecount}[theocount]
\newcounter{examplecount}[section]
\theoremstyle{plain}
\newtheorem{theo}[theocount]{\iflanguage{french}{Théorème}{Theorem}}
\newtheorem{corollaire}[corollairecount]{\iflanguage{french}{Corollaire}{Corollary}}
\newtheorem{lemme}[propcount]{\iflanguage{french}{Lemme}{Lemma}}
\newtheorem{proposition}[propcount]{Proposition}
\newtheoremstyle{properties}
{}%      Space above, empty = `usual value'
{}%      Space below
{\normalsize\normalfont\itshape}% Body font
{}% Indent amount (empty = no indent, \parindent = para indent)
{\normalsize\normalfont\bfseries}% Thm head font
{:}%        Punctuation after thm head
{ }%  Space after thm head: " " = normal interword space;
{}% Thm head spec

\theoremstyle{properties}
\newtheorem{property}[propcount]{\iflanguage{french}{Propriété}{Property}}

\newcommand{\defaultlabelenumi}{\labelenumi}

\theoremstyle{definition}
\newtheorem{definition}[definitioncount]{\iflanguage{french}{Définition}{Definition}}

\theoremstyle{remark}
\newtheorem*{rem}{\iflanguage{french}{Remarque}{Remark}}

\newtheoremstyle{example}
{}%      Space above, empty = `usual value'
{}%      Space below
{}%\footnotesize Body font \sffamily
{}% Indent amount (empty = no indent, \parindent = para indent)
{\normalsize\normalfont}% Thm head font \bfseries
{.}%        Punctuation after thm head
{ }%  Space after thm head: " " = normal interword space;
{}% Thm head spec
\theoremstyle{example}
\newtheorem{exemple}[examplecount]{\iflanguage{french}{Exemple}{Example}}

\newcommand{\E}[1][]{\underset{i}{!} #1}

\mathcode`\?="003F
\mathcode`\!="0021
\begin{document}
\title{Non uniform (hyper/multi)coherence spaces} 

\author{Pierre Boudes\\
 Laboratoire d'Informatique de Paris Nord (UMR
  7030)\\
CNRS/universit\'e Paris nord, institut Galil\'ee \\
99 av. J.-B. Cl\'ement, 93430 Villetaneuse, France\\ \texttt{pierre.boudes@lipn.univ-paris13.fr}
}

%\input{body}
%\setlength{\textfloatsep}{2pt}
%\setlength{\intextsep}{2pt}

%\selectlanguage{english}

\maketitle

\begin{abstract}
In (hyper)coherence semantics, proofs/terms are cliques in
(hyper)graphs. Intuitively, vertices represent results of computations
and the edge relation witnesses the ability of being assembled into a
same piece of data or a same (strongly) stable function, at arrow
types.

In (hyper)coherence semantics, the argument of a (strongly) stable
functional is always a (strongly) stable function. As a consequence,
comparatively to the relational semantics, where there is no edge
relation, some vertices are missing. Recovering these vertices is
essential for the purpose of reconstructing proofs/terms from their
interpretations. It shall also be useful for the comparison with other
semantics, like game semantics.

In~\cite{phaseexp}, Bucciarelli and Ehrhard introduced a so called
\emph{non uniform coherence space semantics} where no vertex is
missing. By constructing the co-free exponential we set a new version
of this last semantics, together with non uniform versions of
hypercoherences and multicoherences, a new semantics where an edge is
a finite multiset. Thanks to the co-free construction, these non
uniform semantics are deterministic in the sense that the intersection
of a clique and of an anti-clique contains at most one vertex, a
result of interaction, and extensionally collapse onto the
corresponding uniform semantics.
\end{abstract}

\tableofcontents

\paragraph{Notations.}
%Multisets on a set $A$ are application from $A$ to $\bbbn$. 
In this paper, multiset will always means \emph{finite} multiset. We
use the notation $[~~]$ for multisets while the notation $\{~~\}$ is,
as usual, for sets. The pairwise union of multisets is denoted by a
$+$ sign and following this notation the generalised union is denoted
by a $\sum$ sign. The neutral element for this operation, the empty
multiset, is denoted by $[]$. If $k\in\bbbn$, $k[a]$ denotes the
multiset $\sum^k_1[a]$.  If $[a_i\mid i\in I]$ is a multiset, its
support is the set $\{a_i\mid i\in I\}$. The cardinality $\card
[a_i\mid i\in I]$ of a multiset $[a_i\mid i\in I]$ is the cardinality
$\card I$ of the set $I$. If $m$ is a multiset we denote by $\supp{m}$
its support. The disjoint sum operation on sets is defined by setting
$A+B=\{1\}\times A\cup \{0\}\times B$. The categorical composition is
denoted by $\comp$.

\section{Introduction}

\subsection{Strong stability and hypercoherences}
Strong stability has been introduced by Bucciarelli and Ehrhard
in~\cite{seqext} for the purpose of giving a purely ``extensional''
definition of sequentiality at all types, that is, a description of
sequential computations which does not involve the atomic description
of each step of interaction of an agent (function, term) with its
environment (argument, or more generally, context), as game semantics
do. The results obtained by Ehrhard in~\cite{relpcf} and later proved
again by Longley~\cite{Longley}, Van~Oosten~\cite{VanOosten} and
Melli\`es~\cite{melliesseqext}, with different methods, showed that
indeed, strong stability corresponds to sequentiality at all
types. Ehrhard established that the strongly stable semantics is the
extensional collapse of the sequential algorithm semantics designed in
the late 70's by Berry and Curien~\cite{cds}.  Unlike the continuous
or stable interpretations of PCF, the sequential algorithm
interpretation (which is now better understood as a deterministic game
semantics) is very ``operational'' in nature: Cartwright, Curien and
Felleisen showed in~\cite{CartwrightCurienFelleisen94} that sequential
algorithms are fully abstract (and fully complete) for the extension
of PCF by a \emph{catch and throw} mechanism.
%  The intuition that strong stability
% is relevant from an operational viewpoint is further supported by
% recent results showing for instance that the strongly stable semantics is
% the extensional collapse of an extension of PCF with states
% (\cite{gsa}). 
In~\cite{Longley}, Longley advocates the claim that
there exists a canonical notion of ``sequential'' functionals of all
types which coincides with the hierarchy of strongly stable functions.
%(in the ``effective'' hierarchy, the situation is more complicated
%however).

This comparison of the strongly stable semantics with more operational
interpretations has been made possible only by the discovery of
\emph{hypercoherences} by Ehrhard~\cite{hyper}. Moreover, the
introduction of these objects simplified the presentation of the
strongly stable semantics and provided a strongly stable
interpretation of (second order) linear logic. A hypercoherence is
very similar to a coherence space~\cite{linearlogic} and consists of a
set, the \emph{web}, together with a coherence relation on this web.
However, in a hypercoherence, the coherence relation is not a binary
relation, but a set of finite subsets of the web containing all
singletons (these sets are said to be coherent). An ``element'' of a
hypercoherence $X$ is then a \emph{clique} of $X$, that is, a subset
of the web of $X$ which has the property that all its finite and
non-empty subsets are coherent.

Hypercoherences are a semantics of linear logic, so they provide an
interpretation of intuitionistic implication which is of the shape
$X\Rightarrow Y=(!X)\multimap Y$ where ``$\multimap$'' is a linear
implication and $!$ is a so called ``exponential''. The basic
operational intuition behind this decomposition is as follows: a
linear map represents a program which uses its argument exactly once,
and an element of $!X$ is obtained essentially by taking an element of
$X$ and making it available as many times as required.

The purely relational semantics is maybe the simplest semantics of
linear logic. In this semantics formul{\ae} are sets and proofs are
relations. The constructions of the relational semantics underly both
the coherence space semantics and the hypercoherence semantics.
Barreiro and Ehrhard traced back the introduction of the relational
semantics as induced by an unpublished remark from van de~Wiele about
the co-freeness of the exponentials in coherence semantics.

The hypercoherence semantics is said to be static as opposed to games
semantics which involve a direct representation of the dynamics of
computation. In game semantics, time is explicit: such semantics
interpret terms by focusing on the history of an atomic interaction
between a player (the program implemented by the term) and an opponent
(the environment).  For instance an interaction inside a function type
$A\to B$ is an interleaving of an interaction querying a piece of $A$
data and an interaction producing a piece of $B$ data.

There is no such reference to time in hypercoherences.
% indeed, the web constructions in the hypercoherence interpretation
% follow the patterns of the purely relational semantics of linear logic.
%In particular 
For instance, the web of a linear function space is the Cartesian
product of the webs.

However the strong relation of hypercoherences with sequentiality
means that the semantics carries an \emph{implicit} representation of time.
% which is missing in the purely relational
% semantics. 

In~\cite{melliesseqext}, Melli\'es investigate the game theoretic
counterpart of this implicit representation of time, by introducing
sequential games in which the coherence relation can be expressed in
game terms.

In a complementary direction, we used an unfolding of hypercoherences
introduced by Ehrhard in~\cite{sp}, to uncover the game structures of
hypercoherences~\cite{pbg}. For the reverse direction, the idea was to
project directly usual games onto hypercoherences by mapping history
of interaction onto results of interaction. But some history of
interaction were not mapped anywhere in the hypercoherence
semantics. Indeed, not only the representations of interaction differ
between games and hypercoherences but these two kinds of semantics do
not agree on what are the possible interaction between terms. More
precisely, one can circumvent the problem by projecting onto the
relational semantics rather than
hypercoherences~\cite{gamesexperiments}. In the relational semantics
the representation of an interaction is the same as in hypercoherences
but their is less (or no) assumption in the relational semantics about
the possibilities of interactions.
 
\subsection{Uniformity}
% TODO rephrase the following sentence
The relational semantics almost consists of the part of the
hypercoherence semantics dealing with webs, except that in hypercoherences
the web of the exponentials depends on the coherence relations. To be
precise the web of $!A$ in the relational semantics is the set of
multisets of elements of the web of $A$ while, in hypercoherences the
web of $!A$ contains only multisets which supports are clique.

The dependence of webs on coherence is what is called uniformity of
the exponentials.  This terminology, mainly used by Ehrhard and
Girard, comes from the fact that in such semantics the context of an
agent behaves uniformly, that is: as if this context is produced by a
single agent. 
% Making the difference between
% the static information which concerns types and which is delivered by
% the web and the dynamic information which concerns terms and which is
% delivered by the coherence relation is hard since uniformity mixes
% them up.
The hypercoherence interpretation of a term omits points relatively to
its relational interpretation and so the hypercoherence semantics
loses information about some parts of the term. The same holds for the
coherence semantics.

Lets take an example (very standard). % very standard example.
The relational semantics of the simply typed term
\begin{equation}
\lambda b^{\bool}.
\Ifthenelse{b}{(\Ifthenelse{b}{\true}{\true})}{%
(\Ifthenelse{b}{\false}{\false})}.\label{eq:2}
\end{equation}
(where $\true$ stands for true and $\false$ for false) is the relation
\[
\{([\true,\true],\true), ([\false,\true],\true),
([\false,\true],\false), ([\false,\false],\false)\}
\]
but its hypercoherence semantics is just $\{([\true,\true],\true),
([\false,\false],\false)\}$.

The hypercoherence semantics of the term trusts its environment and
makes the assumption that the boolean $b$ has one fixed value during
the time of the computation. Of course, this is fair from an
interactive viewpoint since the environment complies with coherence
conditions as programs do. But for the purpose of reconstructing terms
from their semantics, some information is missing. Our example is very
simple, but it is easy to imagine terms where the part missed by the
uniform interpretation contains big sub-routines rather than
constants.

Intuitively, one can think of uniformity as a technique to remove a
particular kind of dead code, as in the example above.  However, it is
worth to remark that what is lost by uniformity in coherence spaces
and hypercoherences can hardly, in general and especially at
functional types, be match with well-identified pieces of syntax
(sub-terms or sub-proofs). First these semantics are not fully
complete, so (a part of) a semantical agent not always corresponds to
(a part of) a syntactical agent.  Second the uniformity restriction
not only takes into account accessibility of branches of code but
also, in a more subtle manner, reachability constraints, in particular
on the copying discipline of pieces of data (see
Example~\ref{ex:another}).

Non uniform static semantics will interpret terms exactly as the
relational semantics does. This will allow us to \emph{combine}
semantics in order to take advantage of their different features. For
instance, we can define a semantics where proofs will be cliques both
in the non uniform coherence space semantics, in the non uniform
hypercoherence semantics and in the non uniform bipartite semantics
we present in this paper.

The uniformity/non uniformity issue in static semantics is to be
related with games where some uniformity conditions were originally
designed for the exponential type : interactions in $!A$ are
\emph{deterministic} (in the sense of games) interleaving of
interactions of $A$, see \cite{ajm}. Recent works in the game
semantics area are more permissive: such conditions (games
determinism) on the semantics of types are postponed to conditions on
the semantics of terms.

\subsection{The former (hyper)coherence semantics}

Providing coherence space or hypercoherence semantics with non
uniform exponentials is not a trivial job. One has to design a
semantics where for instance, one point of the web shall be incoherent
with himself. This must be the case for the point $[\true,\false]$
since the valid term above maps it to an incoherent piece of data
$\{\true, \false\}$.  The situation where two different points are
coherent and incoherent at the same time may also arise. In coherence
spaces this will mean the semantics does not enjoy \emph{determinism}
---we come back with this latter.

However, the main difficulty lies in defining the interpretation of
the exponentials.

A.~Bucciarelli and T.~Ehrhard have designed a general tool for
producing non uniform semantics see~\cite{phaseexp}. As observed by
J.-Y.~Girard in~\cite{ondenot}, to be closer to full completeness for
linear logic, the coherence spaces semantics can be enriched by
indexing each clique on a monoid. To make the story short, by doing
this and thanks to a clever handling of \emph{indexes} (locations),
A.~Bucciarelli and T.~Ehrhard obtained that when this monoid comes
with a \emph{phase space} structure of a certain sort (actually, a
symmetric phase space which is a truth-value semantics of an
\emph{indexed} linear logic) this leads to a denotational semantics of
linear logic. For details see~\cite{phasemall} and~\cite{phaseexp}.
This leaves us with, potentially, an infinity of denotational semantics
of linear logic.  A.~Bruasse-Bac has studied many of them in her PhD.
thesis~\cite{alexphd} among which there is one rejecting the
\emph{Mix} rule. A quite simple phase space produces a former version
of non uniform coherence semantics. According to a suggestion of
Ehrhard and Bucciarelli, by generalizing the construction to all
arities one obtains non uniformity for something sounding like non
uniform hypercoherence semantics. But:
\begin{enumerate}
\item each of these non uniform semantics badly relates with their
  usual (\eg uniform) versions;
\item neither the coherence, nor the hypercoherence non uniform
  semantics are deterministic;
\item Furthermore, we observed that the former non uniform
  hypercoherence semantics misses one important feature of the usual
  hypercoherence semantics : not all the finite cliques of type
  $\bool^n\to\bool$ are \emph{sub-definable} (\ie included in the
  interpretation of a term).  

\item The former non uniform coherence semantics is also a little bit
  puzzling. For instance, in $!\bool$, for any $p, q\in \bbbn$ such
  that $p+q>1$, the points $p[\true]$ and $q[\true]$ are coherent and
  incoherent at the same time (while, once $p\neq q$, they are
  strictly coherent in usual coherence spaces).  So, one can find in
  the former non uniform semantics a semantical agent mixing the
  term~(\ref{eq:2}), querying two times its argument, and the term
  $\lambda b. \true$, which does not use its argument.  (This is not
  the case in the former hypercoherence semantics).
\end{enumerate}

\subsection{Contribution}

The present work is an extended version of our previous communication
at the CTCS conference~\cite{nuhc}. Some parts were also presented in
our PhD.

Our starting point was the former non uniform (hyper)coherence
semantics and we mainly focused on hypercoherences. We observed that:
\begin{enumerate}
\item contrarily to what happens in the usual hypercoherence
  semantics, there is at least one clique $f$ of type
  $\bool^n\to\bool$ which is not included in the interpretation of any
  term. The clique $f$ is a variant of one originally designed by
  Berry to reveal the same failure in coherence
  spaces\footnote{Berry's example is often called the
    \emph{Gustave's function} which is named after a private joke
    about the huge number of french scientists whose first name is
    G\'erard, among which Berry.}. What is important to notice is that
  $f$ does not contain points related to non uniformity, such as
  $[\true,\false]$. Hence the set $f$ can be presented to the usual
  (multiset based) hypercoherence semantics which successfully refutes
  it.
\item Many others definitions of the interpretation of exponentials
  are possible.
\end{enumerate}

Among all the possibilities for the interpretation of exponentials we
found the \emph{co-free} exponential (think of it as to be an
\emph{infinite} tensor product). This has led to a more satisfactory
setting both for coherence spaces and hypercoherences. We also
introduced a new coherence like semantics, multicoherences.

\begin{enumerate}
\item The co-free exponential is maximal in a sense we make precise in
  Corollary~\ref{cor:max} and which basically means that any clique of
  type $\bool^n\to\bool$ would also be a clique with others variants
  of the exponentials. The bad news is that there still exists a
  clique of type $\bool^n\to\bool$ which is not included in the
  interpretation of any (sequential) term. But, in hypercoherences
  (and multicoherences), such cliques necessarily contain points
  coming from non uniformity. So, this phenomenon is now constrained to
  the non uniform web and disappears when restricting to the uniform
  web.

  % Now the good news
\item Our non uniform semantics are \emph{deterministic} in the sense
  that the intersection of a clique (let say of type $A$) and an
  anti-clique (a clique of dual type $A\orth$) contains at most one
  point, like in the uniform semantics. We call \emph{results of
    interactions}, the points at the intersection of a clique and an
  anti-clique. Not all points can be results of interactions. For
  instance $[\true, \false]\in !\bool$ is not, since it is strictly
  incoherent with itself (and so does not belong to any clique of type
  $!\bool$).

\item For both coherence spaces, hypercoherences and multicoherences,
  we derive the usual uniform interpretation of types and terms by a
  straight set restriction of the web to results of interactions (both
  at the the level of types and at the level of proofs). More
  precisely, the restriction gives the multiset based uniform
  interpretation (the web of $!A$ is made of multisets which supports
  are cliques). Note that is was already known that the restriction of
  the relational interpretation of a proof to the web of the coherence
  space interpretation of a formula is the coherence space
  interpretation of the proof~\cite{phd:Lorenzo}. The novelty
  introduced here is mainly that the uniform web is characterized by
  the coherence relation.

\item In each case, the non uniform and the multiset based uniform
  semantics are extensionally equivalent : they have the same
  extensional collapse. As already
  known~\cite{NunoCollapse,melliescompcollapse} for the multiset based
  semantics, this extensional collapse is the set based uniform
  semantics.

\item The existence of deterministic non uniform semantics implies an
  unexpected property. Consider an extension of linear logic with new
  rules, typically the daemon of Girard's Ludics~\cite{locus}, for
  which the semantics is still valid and such that $A$ and $A\orth$
  are both provable. Then a cut between a proof of $A$ and a proof of
  $A\orth$ induces an interaction which involves at most one point in
  the relational semantics.  This unexpected result would have been
  hard to prove without introducing non uniform (hyper/multi)coherence
  semantics. It may prove useful with other semantics based on the
  relational semantics. For instance, in Ehrhard's finiteness
  spaces~\cite{finiteness}, points of interaction are equipped with
  multiplicities. Since there is at most one point for each
  interaction, one can use its multiplicity to do some quantitative
  analysis of proofs' interaction.

\item The multicoherence semantics aroses as the general case in our
  approach of non uniformity. In fact, we derive the non uniform
  hypercoherence semantics from multicoherences. The difference with
  hypercoherences is that the coherence relation is made of multisets
  rather than sets. As for uniform hypercoherences, at first order
  simple types, each finite clique in the uniform multicoherence
  semantics is sub-definable. But contrarily to hypercoherences, each
  clique of the multicoherence semantics is a clique in coherence
  spaces. At a functional type, there exists sets which are cliques in
  the hypercoherence semantics but which are not cliques in the
  multicoherence semantics (even in the set based uniform case). Since
  the set based uniform multicoherence semantics is extensional, there
  is at least two extensionally different semantics of higher order:
  multicoherences and hypercoherences.

% \fbox{But but but...} est-ce
%   qu'une preuve est interprétée par le même ensemble de points dans
%   les hypercoherences et dans les multicoherences ? Sinon est-ce qu'on
%   peut montrer que la restriction aux preuves est extensionnellement
%   égale (si j'arrive à donner du sens à ça).

%\item a little bit stronger than just determinism.... In usual (hyper)coherence semantics, the compos
\end{enumerate}

In Table~\ref{table:summary}, we summarize the principal variants of
coherence based semantics, with our new ones. Note that there are two
axis where one can chose between sets and multisets: either for the
coherence relation (the power of the coherence) or for the web of the
exponentials. A third one, which do not appear in that paper, is the
shift from sets to multisets for cliques, as in finiteness
  spaces~\cite{finiteness}.

The (non uniform) bipartite semantics of linear logic we present in
Section~\ref{sec:bip} comes originally from a simple remark about the
relational semantics. This remark states that one can set polarities
(positive/negative) on points of the web in such a manner that the
orthogonal exchanges polarities and that every proof is interpreted by
a set of positive points. This gives a kind of coherence spaces
semantics where the coherence relation is of arity one. The semantics one
obtains is non uniform. We discovered that this bipartite semantics
also admits a uniform version where every proof of a \emph{why not}
formula is interpreted by the empty set. Besides this radical lapse of
memory of points in proofs interpretation, the uniform bipartite semantics
(as the non uniform bipartite semantics) is equivalent to the relational
semantics on simple types (proofs and types interpretations are the same
and these three semantics are extensionally equivalent).

\begin{table*}
\centering
  \noindent
\footnotesize
  \caption{A summary}
  \begin{tabular}{|l|c|c|c|c|c|}
    \hline
    \bf Semantics : & relational & coherence spaces & hypercoherences &
        multicoherences & bipartite \\
%    \hline
    \bf Coherence : & empty & pairs & sets & multisets
    & singletons\\
%    \hline
    \bf Category : & \Rel & \NCohK[\{2\}] & \NHC & \NCohK[\bigK] & \Bip\\
\hline
    \multicolumn{6}{|l|}{\bf Web of the exponentials (variants)} \\
    \hline
    multisets (co-free)  & yes & new version & new & new & new\\
    multisets uniform & no & yes & yes & new & new\\
    sets uniform & no & yes & yes & new & new \\
    \hline
  \end{tabular} 
\label{table:summary}
\end{table*}

In a recent work~\cite{pagani06mscs} Michele Pagani showed that there
is a syntactical counterpart, \emph{visible acyclicity}, to non
uniform coherence : our non uniform coherence space semantics
corresponds to a relaxation of the correctness criterion of linear
logic proof-nets, (a graphical presentation of proofs). Finding a
similar correspondence for non uniform hypercoherences or
multicoherences would certainly be interesting.

\subsection{Contents}
The next section (Section~\ref{sec:prelim}) is devoted to recalling briefly some useful definitions and properties we deal with. The only novelty is the introduction of a convenient framework, \Pcoh spaces, to deal with various static semantics.

In Section~\ref{sec:bip}, we present the bipartite semantics and its
uniform version, and we compare them. We stress that the bipartite
semantics are just here as a peculiar example of uniform/non
uniform setting, but do not give an example of the general uniform/non
uniform construction we use further.

Section~\ref{sec:core} and Section~\ref{sec:relunifnonunif} form the
core of the paper, where we study the exponential coming from indexed
linear logic, and develop their co-free version. This part is an
extended version of our conference paper on non-uniform
hypercoherences~\cite{nuhc}. In this part, we present our results by
mostly following the chronological order of their discovery.

We start with the presentation of \Kcoh spaces, a denotational
semantics coming from indexed linear logic and aimed to be a
generalization of (hyper)coherence spaces. The semantics is
parametrized by a set $K$ which encompasses (a kind of) coherence
spaces $K=\{2\}$ and (a kind of) hypercoherences $K=\bigK$. We then
point out the definability problem at first order.

We further introduce the co-free exponentials
(Subsection~\ref{sec:free}). We show that the co-free exponentials
provide the best solution one can expect (Corollary~\ref{cor:max}).
We show that there are still some definibility problems
(Corollary~\ref{cor:nodef}) but only due to non-uniformity.

We then establish (Subsection~\ref{sec:determinism}) that the \Kcoh
semantics equipped with the new exponentials gives a deterministic
semantics. In fact, the semantics satisfies a property stronger than
just determinism. This allows the introduction of a web restriction
operation, the neutral functor (Subsection~\ref{sec:neutralweb})
giving the uniform version of the \Kcoh semantics
(Subsection~\ref{sec:uniform-k-coherence}).

The uniform semantics obtained for $K=\bigK$ is not hypercoherences
but a new one, multicoherences (Subsection~\ref{sec:multicoh}).
Hypercoherences and non-uniform hypercoherences are obtained by an
operation forgetting multiplicities in the coherence relation
(Subsection~\ref{sec:hypercoh}).

We close this part by a study which relates the extensional collapses
of the various semantics (Subsection~\ref{sec:collapses}).

In the concluding section, we adopt an interactive viewpoint \`a la
Girard to discuss the implications of the existence of deterministic
non-uniform semantics (Subsection~\ref{sec:staticinteractivity}).

\section{Preliminaries}
\label{sec:prelim}

\subsection{Extensional collapse}

%Extensional collapse is the
%quotient by extensional partial equivalence relations (PERs).
Extensional partial equivalence relations were first introduced by
Kreisel in the fifties to deal with higher order partial recursive
functions. An extensional PER is meant to relate two algorithms when
they implement the same function. Higher order is
responsible for the partiality of the equivalence relation.

Simple types are types of the simply typed lambda calculus enriched
with basis types in order to form a type system for PCF.
% calculus
%(programming computable functions, of D.~Scott and G.~Plotkin. 
They are given by the following grammar:
\begin{align*}
  \sigma,\tau &:=\iota\mid\sigma\to\tau\quad\text{(simple types)}
\end{align*}
where $\iota$ stands for basis types, typically a boolean type $\bool$
or/and a natural number type $\nat$. A product type $\sigma\times
\sigma'$ can also be introduced but we won't bother with this type
constructor since it can be obtained by curryfication.
\newcommand{\M}{{\cal M}}

Let $\cal M$ be a categorical semantics of linear logic, let $\M(A)$
denotes the interpretation of a type $A$ and let us call semantical
agents of type $A$ the elements of the semantics used to interpret
proofs of $A$, that is morphisms from $1$ to $\M(A)$. Suppose an
interpretation of basis types is given in $\M$ (usually $\bool$ is
interpreted as the space $1\oplus 1$ and $\nat$ is interpreted as an
$\omega$-infinite \emph{plus} of $1$). Then we extend this
intepretation of basis types to all simple types by setting
$\M(\sigma\to\tau)=!{\cal M}(\sigma)\loli {\cal M}(\tau)$. So, the
function type corresponds to the object of morphisms in the co-Kleisli
category. If $f$ and $x$ are semantical agents of respective types
$\sigma\to\tau$ and $\sigma$ then we \emph{apply} $f$ to $x$ by
composing in the co-Kleisli category to form a semantical agent $f(x)$
of type $\tau$. For each simple type $\sigma$, an extensional PER
$\sim_{\sigma}$ is defined on semantical agents of type $\sigma$ by
chosing the equality on basis types and by setting:
\[
f\sim_{\sigma\to\tau} g\text{ iff if }x\sim_{\sigma} y\text{ then
 }f(x)\sim_{\tau} g(y).
\]

The \emph{extensional collapse} of the semantics is the set of quotients
by extensional PERs of the interpretations of simples types equipped
with the following notion of application. If $\overline{f}$ is a class
of semantical agents of type $\sigma\to\tau$ (functions) and if
$\overline{x}$ is a class of semantical agents of type $\tau$
(arguments) the application of $\overline{f}$ to $\overline{x}$ is
defined by setting $\overline{f}(\overline{x})=\overline{f(x)}$.

\subsection{Power coherence spaces}
We introduce a general notion which will provide us with a very
convenient language for describing the various semantics we deal with. A
\emph{power} is simply a functor from the category of sets and
inclusions to itself. Typical powers relevant to our purpose are:
\begin{itemize}
\item The empty power defined by $E \mapsto \emptyset$. This power will
  simply be denoted $\emptyset$. It can be used to present the
  relational semantics in terms of power coherence spaces;
\item The identity power, $\id[]$, which will be used for dealing with
  the bipartite relational semantics of Section~~\ref{sec:bip};
\item The non-empty finite sets power $\Pfinast$ which maps each set
  to the set of its finite non-empty subsets. The power $\Pfinast$
  will be used for dealing with hypercoherences;
\item Given a subset $K$ of $\bbbn\setminus\{0,1\}$, the power
  $\MultK$ which maps a set $E$ to the set of all finite multisets
  over $E$ whose cardinality belongs to $K$. The power $\MultK[\{2\}]$
  will be used for dealing with coherence spaces. The choice of this
  power follows the suggestion made at the end of \cite{phaseexp} for
  the purpose of building non uniform coherence or hypercoherence like
  semantics.
\end{itemize}

\begin{definition}\label{def:Pcoh}  
  Let $P$ be a power. A \emph{\Pcoh space} $X$ is a triple
  $(|X|,\Rcoh,\Rincoh)$ where $|X|$ is an at most countable set, the
  \bindex{web} of $X$, and where $\Rcoh$ and $\Rincoh$ are subsets of
  $P(|X|)$ such that $\Rcoh\cup\Rincoh=P(|X|)$. 
  The set $\Rcoh$ is called the \emph{coherence} and the set $\Rincoh$
  is called the \emph{incoherence}.  The intersection of $\Rcoh$ and
  $\Rincoh$ is called the \emph{neutrality}. Notation: $\Rneutre$. The
  strict coherence $\Rcohs$ of $X$ is the complementary set of
  $\Rincoh$ with respect to $P(|X|)$ and the strict incoherence
  $\Rincohs$ is the complementary of $\Rcoh$.  
\end{definition}

Clearly, one can define a \Pcoh space $X$ by specifying two
well chosen sets among $\Rcoh$, $\Rincoh$, $\Rneutre$, $\Rcohs$ and
$\Rincohs$ subject to obvious constraints (for instance, one must have
$\Rneutre\subseteq\Rcoh$, $\Rincohs\cap\Rcohs=\emptyset$\dots).

\begin{definition}
  \label{def:pcohorth}
  The orthogonal, $X\orth$, of a \Pcoh space $X=(|X|,
  \Rcoh, \Rincoh)$ is the \Pcoh space $(|X|, \Rincoh,
  \Rcoh)$. (Orthogonality exchanges coherence and incoherence).
\end{definition}

\begin{definition}\label{def:pcohclique}
  Let $X$ be a \Pcoh space. A \bindex{clique} of $X$ is a subset $x$
  of $|X|$ such that $P(x)\subseteq\Rcoh$. We denote by $\Cl(X)$ the
  set of all cliques of $X$. An \bindex{anti-clique} of $X$ is a
  clique of $X\orth$. If for each clique $x$ and each anti-clique $y$
  the intersection of $x$ and $y$ contains at most one element then
  \Pcoh space $X$ is \bindex{deterministic}.
\end{definition}

\begin{definition}
  A \Pcoh space $X$ is \bindex{reflexive} if neutrality corresponds to
  equality in the sense that:
  \begin{gather}
    \Rneutre = \cup_{a\in |X|} P(\{a\}).\label{eq:reflex}
  \end{gather}
  A \Pcoh space $X$ is \bindex{weakly reflexive} if
  \begin{gather}
    \Rneutre \subseteq \cup_{a\in |X|} P(\{a\}).\label{eq:weakreflex}
  \end{gather}
\end{definition}

One can define a reflexive \Pcoh space by specifying only $\Rcoh$ (or
$\Rincoh$, or $\Rcohs$, or $\Rincohs$).

\begin{proposition}
 \label{prop:weakreflexdetermin}
  If the power is strictly monotone and preserves disjointness of
  sets, then weak reflexivity implies determinism.
\end{proposition}

\begin{proof}
  Lets take a \Pcoh space $X$, and a clique and an anti-clique with at
  least two points, $a$ and $b$ at their intersection. Then $\{a, b\}$
  is both a clique and an anti-clique.  Thus $P(\{a,b\})\subseteq
  \Rcoh\cap\Rincoh$. By strict monotonicity and preservation of
  disjointness $P(\{a, b\})\not\subset \cup_{c\in |X|} P(\{c\})$ which
  contradicts weak reflexivity.
\end{proof}

But weak reflexivity is in general stronger than just determinism.
For instance, in a $\Pfinast$-space $X$ one can find a set $\{a, b,
c\}\in\Pfinast(|X|)$ which is both coherent and incoherent (so the
space is not weakly reflexive) and still have determinism (take for
instance $\Rcohs = \{\{a,b\},\{a,c\}, \{b, c\}\}$ and
$\Rincohs=\emptyset$).

Weak reflexivity and determinism are equivalent in
$\MultK[\{2\}]$-spaces.

\subsection{Relational semantics}

We recall briefly the interpretation of linear logic in the category
\Rel of sets and relations.

Let us recall that the composition is given by:
\[f\comp g=\{(a,c)\mid
\exists b, (a,b)\in f\et (b,c)\in g\}\] and that identities are given
by: \[\id[X]=\{(a,a)\mid a\in
|X|\}.\]

\begin{description}
\item[Formulae.]  A formula $A$ is interpreted by a set $|A|$ defined
  inductively as follows: $|0|=|\top|=\emptyset$,
  $|1|=|\bot|=\{\ast\}$, $|A\orth|=|A|$, $|A\oplus B|=|A\& B|=|A|+
  |B|$, $|A\otimes B|=|A\Par B|=|A|\times |B|$ and
  $|!A|=|?A|=\Mfin(|A|)$ where $\Mfin(E)$ is the set of finite
  multisets on $E$.
  
\item[Sequents.] We use the right-sided presentation of the linear
  logic sequent calculus. Up to associativity and commutativity of the
  Cartesian product, the ``comma'' of sequents is safely interpreted
  as a \emph{par} \ie by setting ${|\vdash
  A_1,\ldots,A_n|}=|A_1\Par\ldots\Par A_n|$ which is equal to
  $|A_1|\times\ldots\times |A_n|$.
  
\item[Proofs.] The interpretation of a proof of a sequent
  $\vdash\Gamma$ is a subset of $|\vdash\Gamma|$ defined inductively
  on the proof, by cases on the last rule, as shown below.
\end{description}

It is well-known that this interpretation is a denotational semantics
of linear logic (that is: two proofs of a given sequent have the same
interpretation as soon as they are equivalent up to cut-elimination).

\medskip
\begin{center}
\textbf{Identity group}\hfill~\nopagebreak\\ %<-- ne pas casser la ligne
  \AXR{A,A \orth:\{(a,a)\mid a\in |A|\}} \tabDP
  \AXC{$\vdash\Gamma,A:f$} \AXC{$\vdash\Delta,A\orth:g$}
  \CUTR{\Gamma,\Delta: \{(\gamma,\delta)\mid\exists a, (\gamma,a)\in
    f\wedge (\delta,a)\in g\}} \tabDP
\end{center}
\medskip
\begin{center}
  \textbf{Additives}\hfill~\nopagebreak\\
  \TOPR{\Gamma,\top: \emptyset} \tabDP\qquad \AXC{$\vdash\Gamma,A:f$}
  \AXC{$\vdash\Gamma,B:g$} \AVECR{\Gamma,A\& B: f\uplus g} \tabDP \qquad
  \AXC{$\vdash\Gamma,A:f$} \PLUSR{\Gamma,A\oplus B: f} \tabDP
\end{center}
\medskip
\begin{center}
  \textbf{Multiplicatives}\hfill~\nopagebreak\\
  \AXC{$\vdash\Gamma:f$} \BOTR{\vdash\Gamma,\bot:
    f\times\{\ast_\bot\}} \tabDP\qquad
  % \AXC{}
  \altUNR{\vdash 1: \{\ast_1\}} \tabDP
  \\
  \AXC{$\vdash\Gamma,A,B:f$}
  \PARR{\vdash\Gamma,A\Par B: f}
  \tabDP\hfill \AXC{$\vdash\Gamma,A:f$} \AXC{$\vdash\Delta,B:g$}
  \TENSR{\vdash\Gamma,\Delta, A\otimes B: \{(\gamma,\delta,(a,b))\mid
    (\gamma,a)\in f, (\delta,b)\in g\}} \tabDP
\end{center}
\medskip
\begin{center}
  \textbf{Exponentials}\hfill~\nopagebreak\\
  \label{relrules:exp}
  \AXC{$\vdash ?A_1,\ldots,?A_n,A:f$}
  \PROMR{?A_1,\ldots,?A_n,!A:f^\dag} \tabDP \quad
  \AXC{$\vdash\Gamma,?A,?A:f$} \CONTR{\Gamma,?A:
    \{(\gamma,\mu_1+\mu_2)\mid (\gamma,\mu_1,\mu_2)\in f\}} \tabDP
  \\
  \AXC{$\vdash\Gamma:f$} \AFFR{\Gamma,?A: \{(\gamma,[])\mid
    (\gamma)\in f\}} \tabDP\qquad \AXC{$\vdash\Gamma,A:f$}
  \DERR{\Gamma,?A: \{(\gamma,[a])\mid (\gamma,a)\in f\}} \tabDP
  \\
  Where $f^\dag$ is equal to :\hfill~
  \[
  \{(\sum_{j\in J} \mu_1^j,\ldots,\sum_{j\in J} \mu_n^j, [a_j\mid j\in
  J])\mid [(\mu_1^j,\ldots,\mu_n^j, a_j)\mid j\in J]\in \Mfin(f)\}.\]
\end{center}
\medskip The relational semantics is actually a categorical semantics
of linear logic, though we shall not recall its categorical structure
in details. The new Seely categorical semantics
axiomatic~\cite{whatillcatmod} is appropriate for dealing with the
relational semantics and we will use this axiomatic for further
semantics.  Exponentials are given by a comonad structure
$(!,\der,\dig)$. We just recall this structure. The endofunctor $!$ of
\Rel is defined by $!E=\Mfin(E)$ and \[ !f=\{([a_i\mid i\in
I],[b_i\mid i\in I])\mid [(a_i,b_i)\mid i\in I]\in \Mfin(f)\}.
\] 
The natural transformations $\der:!\dot{\to}\id$ and
$\dig:!\dot{\to}!!$ are defined by setting:
\begin{align*}
  \der[E]&=\{([a],a)\mid a\in E\}\\
  \dig[E]&=\{(\sum_{i\in I} \mu_i,[\mu_i\mid i\in I])\mid [\mu_i\mid
  i\in I]\in !!E\}.
\end{align*}

\subsection{Coherence spaces}

We briefly recall the coherence spaces semantics of linear
logic~\cite{linearlogic}. 

A coherence space is a reflexive $\MultK[\{2\}]$-coherence space.

We define directly the connectives of linear logic on coherence spaces
(rather than defining by induction the interpretation of
formulae). The web of multiplicatives and additive is the same as in
the relational semantics. Coherence is defined as follows.  One has
$\Rcoh[X\oplus Y] = \Rcoh[X]\uplus\Rcoh[Y]$ and $[(a, b), (a',
b')]\in\Rcoh[X\otimes Y]$ iff $[a, a']\in\Rcoh[X]$ and $[b,
b']\in\Rcoh[Y]$.

Morphisms from $X$ to $Y$ in the corresponding (linear) category are just
cliques of $X\loli Y = X\orth \Par Y$.

The interpretation of proofs in coherence spaces only differs from the
relational semantics on exponential rules. Let us recall that the
interpretation of a proof is just a subset of the web of the
interpretation of the conclusion sequent and that this is a corollary of the
soundness of the semantics that every such set is a clique in the
corresponding space.

There are two variants for the web of the exponentials : set based and
multiset based. In the multiset based semantics, the web of $!X$ is
the set of multisets which supports are cliques in $X$:
\begin{gather}
  |!X| = \{\mu\in\Mfin(|X|)\mid \supp{\mu}\in\Cl(X)\} \label{eq:webofcoursecoh}
\end{gather}

Two elements $\mu$ and $\nu$ of the web of $!X$ are coherent iff the
support of $\mu+\nu$ is a clique of $X$.

The interpretation of exponential rules is the same as in the
relational semantics but restricted to the web of the exponentials.
Two rules have their interpretation modified by this restriction. In
the contraction rule, the support of $\mu_1+\mu_2$ has to be an
anti-clique of (the space interpreting) $A$. In the promotion rule :
(i) the support of $\sum_{j\in J} \mu^j_i$ has to be an anti-clique of
$A_i$, for each $j$; (ii) and the support of $[a_j\mid j\in J]$ has to
be a clique of $A$. In fact, one can easily verify that (i) implies
(ii) so the only condition to check is (i). This will also be the case
in hypercoherences.

In the set based semantics, the web of $!X$ is the set of finite
cliques of $X$. The interpretation of exponentials follows the last
pattern but with sets and unions instead of mulitsets and sums.

\begin{exemple}\label{ex:unexdepreuve}
  In the introduction we gave unformaly the relational interpretation
  and the uniform coherence space interpretation of a term. The proof
  in Figure~\ref{fig:exsemunifpreuve} is a linear logic version of
  this term annotated by the relational interpretation of its
  sub-proofs. In this example we have denoted $(1,\ast)$ by $\true$
  and $(2,\ast)$ by $\false$. The point which is forgotten by the
  coherence space semantics of this proof and the points from which it
  comes from are printed with a line through text.
\end{exemple}
\begin{figure}
\[
\AXL{\rnode{tta}{\bot,1:\{(\ast,\ast)}\}}
\AXL[]{\rnode{fta}{\bot,1:\{\mathsout{(\ast,\ast)}\}}}
\AVECL{\bot\with\bot,1:\{(\true,\ast),\rnode{ftb}{\mathsout{(\false,\ast)}}\}}
\BOTL{\bot\with\bot,1,\bot:\{(\true,\ast,\ast),%
\rnode{ftc}{\mathsout{(\false,\ast,\ast)}}\}}
\AXL[]{\rnode{tfa}{\bot,1:\{\mathsout{(\ast,\ast)}\}}}
\AXL[]{\rnode{ffa}{\bot,1:\{(\ast,\ast)\}}}
\AVECL[]{\bot\with\bot,1:\{\mathsout{(\true,\ast)},(\false,\ast)\}}
\BOTL[]{\bot\with\bot,1,\bot:\{\mathsout{(\true,\ast,\ast)},(\false,\ast,\ast)\}}
\AVECL{\bot\with\bot,1,\bot\with\bot:\{%
(\true,\ast,\true),\rnode{ftd}{\mathsout{(\false,\ast,\true)}},%
\mathsout{(\true,\ast,\false)},(\false,\ast,\false)%
\}}
\DERL{?(\bot\with\bot),1,\bot\with\bot:\{%
([\true],\ast,\true),\rnode{fte}{\mathsout{([\false],\ast,\true)}},%
\mathsout{([\true],\ast,\false)},([\false],\ast,\false)%
\}}
\DERL{?(\bot\with\bot),1,?(\bot\with\bot):\{%
([\true],\ast,[\true]),\rnode{ftf}{\mathsout{([\false],\ast,[\true])}},%
\mathsout{([\true],\ast,[\false])},([\false],\ast,[\false])%
\}}
\CONTL{?(\bot\with\bot),1:\{%
\rnode{ttg}{([\true,\true],\ast)},%
\rnode{ftg}{\mathsout{([\false,\true],\ast)}},%
%\rnode{tfg}\mathsout{([\true,\false],\ast)},%
\rnode{ffg}{([\false,\false],\ast)}%
\}}%
\DP
{%
%%  \psset{linecolor=gray,linewidth=1.5pt,nodesep=1pt,arrows=->}%
%%  \nccurve[angleA=300,angleB=120]{tta}{ttg}%
%%  \nccurve[linestyle=dashed,angleA=270,angleB=90]{fta}{ftg}%
%%  \nccurve[linestyle=dashed,angleA=270,angleB=90]{tfa}{ftg}%
%%  \nccurve[angleA=230,angleB=60]{ffa}{ffg}%
%%  \ncline[linecolor=lightgray,linewidth=2pt,nodesep=1pt]{ftf}{fte}%
%%  \ncline[linecolor=lightgray,linewidth=2pt,nodesep=1pt]{fte}{ftd}%
%%  \ncline[linecolor=lightgray,linewidth=2pt,nodesep=1pt]{ftd}{ftc}%
%%  \ncline[linecolor=lightgray,linewidth=2pt,nodesep=1pt]{ftc}{ftb}%
%%  \ncline[linecolor=lightgray,linewidth=2pt,nodesep=1pt]{ftb}{fta}%
}
\]    
  \caption{linear logic proof of Example~\ref{ex:unexdepreuve}}
  \label{fig:exsemunifpreuve}
\end{figure}

\begin{exemple}
\label{ex:another}
Another example of the action of uniformity concerns restrictions on
the number of times an argument will be copied. This is sligthly more
subtle than just removing pieces of dead code. 

Consider the following proof $\pi_1$ :
\begin{gather*}
  \UNR
  \PROMR{!1.}
  \DP
\end{gather*}
which intuitive meaning is that we make $1$ available \emph{ad
  libidum}. If we add a dereliction as last rule, the proof $\pi_1$
intuitively becomes a program taking as argument another program
requiring an arbitrary number of copies of $1$. For instance, the two
points $([[\ast,\ast]])$ and $([[\ast]])$ in the interpretation of
$\pi_1$ will corresponds to a required number of copies of,
respectively, $2$ and $1$.

Now consider two copies of $\pi_1$ which we assemble into a unique
proof by a combination of bottom and tensor introductions (we could
also have used the \emph{mix} rule if available). We contract the two
copies of $?!1$. The resulting proof $\pi_2$ is shown in
Figure~\ref{fig:another}.
\begin{figure}
  \centering
  \begin{gather*}
    \UNR \PROMR{!1} \DERR{?!1} \BOTR{?!1, \bot} \UNR \PROMR{!1}
    \DERR{?!1} \BOTR{?!1, \bot} \TENSR{?!1, ?!1, \bot\otimes\bot}
    \CONTR{?!1, \bot\otimes\bot} \DP
  \end{gather*}
  \caption{Proof of Example~\ref{ex:another}}
\label{fig:another}
\end{figure}
In coherence spaces, the uniformity restriction forces the two copies
of $\pi_1$ to receive each a program asking for $1$ the \emph{same
  number of times}. For instance, in coherence spaces: $([[\ast,
\ast], [\ast, \ast]], (\ast,\ast))$ is in the interpretation of
$\pi_2$ but $([[\ast], [\ast, \ast]], (\ast,\ast))$ is not; while, in
the relational semantics, both of these points are in the
interpretation. 
\end{exemple}

\subsection{Hypercoherences}

An hypercoherence is a reflexive $\Pfinast$-coherence space.  The
interpretation of linear logic in hypercoherences~\cite{hyper} follows
exactly the same pattern as for coherence spaces.  We just stress a
few points.

In a tensor one has $x = \{(a_1, b_1), \ldots, (a_n,
  b_n)\}\in\Rcoh[X\times Y]$ iff $\pi_1 x = \{a_1, \ldots,
  a_n\}\in\Rcoh[X]$ and $\pi_2 x = \{b_1, \ldots, b_n\}\in\Rcoh[Y]$. In a
\emph{with}, the dual of a \emph{plus}, $\Rincoh[A\with B] =
\Rincoh[A]\uplus\Rincoh[B]$, hence for every $x\in\Pfinast(|A|+|B|)$,
if $x$ intersects both $|A|$ and $|B|$ then $x\in\Rcoh[A\with B]$ (and
conversely).

The two variants for the exponentials, set based and multiset based,
are also presents. 

The coherence in $!X$ is defined using a notion of section.  If
$U=\{x_i\mid i\in I\}$ is a finite set of finite sets or of multisets
we say that $s$ is a \bindex{section} of $U$ and write $s\sect U$ when
for each $i\in I$ there exists $a_i\in s$ such that $a_i\in x_i$ and
$s\subseteq \cup_{i\in I}x_i$. A non empty finite subset $U$ of $|!X|$
is coherent in $!X$ iff each section of $U$ is coherent in $X$.

Of course, since the notion of coherence differs between coherence
spaces and hypercoherence, the notion of cliques and so, because of
uniformity, the interpretation of proofs also differ.

Note that the coherence in hypercoherence may have \emph{holes} : in
general, one can have $x\in\Rcoh$ and $y\subset x$ without having
$y\in\Rcoh$.

\begin{property}
\label{property:cohdeterminism}
  In hypercoherence and coherence space semantics, the intersection
  of a clique and of an anti-clique contains at most one point
  (determinism). But, moreover, in these two semantics, if $(a,c)\in
  f\comp g$ for $f:A\to B$ and $g: B\to C$ then there is only one $b$
  such that $(a,b)\in f$ and $(b,c)\in g$.
\end{property}

Let us recall that one cannot equip the relational semantics with a
set based exponential ($|!X|=\Pfin(|X|)$) similar to the one of
coherence spaces and hypercoherences. This will not give a sound
interpretation. Consider for instance the diagram setting the
naturality of dereliction

\[
\xymatrix{
  {!X}\ar[d]_{!f}\ar[r]^{\der[X]} 
   & {X}\ar[d]^{f}\\
  {!Y}\ar[r]_{\der[Y]} & {Y}
}
\]
In the particular case where $f = \{(a,b), (a',b)\}$, one will have
$(\{a, a'\}, \{b\})\in !f$ so $(\{a, a'\}, b)\in !f\comp \der[Y]$ but
$(\{a, a'\}, b)\notin \der[X]\comp f$. Hence the diagram won't
commute. Saying it in a category free manner, with such a set based
exponentials, the elimination of a cut between a promotion and a
dereliction won't, in general, leave the interpretation invariant.

% Both coherence and hypercoherence semantics admits the \emph{mix}
% rule

\section{Bipartite uniform and non uniform relational semantics}
\label{sec:bip}
In this section, we introduce a simple semantics of linear logic,
based on the relational semantics and $\id[]$-coherence spaces, and which
admits both a non uniform version and a uniform version. We use these
two semantics to demonstrate that uniform semantics can lose a lot of
information about terms (proofs) they interpret.

\begin{definition}
  A \bindex{bipartite space} is just a $\id[]$-coherence space
  $(|X|,\Rcoh,\Rincoh)$ where $\Rneutre$ is empty. So $\Rcoh$,
  $\Rincoh$ is a bipartition of $|X|$ and every bipartition of $|X|$
  defines a bipartite space.
\end{definition}

We further call positive web, denoted $\tramePos{X}$, the coherence of
$X$ and negative web, denoted $\trameNeg{X}$, the incoherence of
$X$. The elements of $\tramePos{X}$ (resp. $\trameNeg{X}$) are the
positive (resp. negative) points of $X$.

Following the general definition (Def.~\ref{def:pcohclique}) in our
particular case, a clique is just a set of positive points of the web.

The category \Bip has bipartite spaces as objets and for morphisms
between bipartite spaces $X$ and $Y$, the relations which are cliques
in the bipartite space $X\loli Y$ defined below. Identities and
composition are those of \Rel. We now describe the non uniform
bipartite semantics of linear logic.

On formul{\ae} the logical connectives are interpreted as follows:

\begin{itemize}
\item linear negation is given, as in the general \Pcoh space case, by
  the exchange of coherence and incoherence, so
  $X\orth=(|X|,\trameNeg{X},\tramePos{X})$;
\item both additives constants $0$ and $\top$ are equal to
  $(\emptyset,\emptyset,\emptyset)$ ;
\item the bipartite space $1$ is equal to
  $(\{\ast\},\{\ast\},\emptyset)$ and, so the bipartite space $\bot$
  is equal to $(\{\ast\},\emptyset,\{\ast\})$;
\item the \emph{with} is given by $X\with
  Y=(|X|+|Y|,\tramePos{X}+\tramePos{Y},\trameNeg{X}+\trameNeg{Y})$ and
  the \emph{plus} is given by $X\oplus Y=(X\orth\with Y\orth)\orth$
  which is the same bipartite space as $X\with Y$;
\item the tensor of $X$ and $Y$, $X\otimes Y$, is the bipartite space
  $|X|\times |Y|$ whose positive web $\tramePos{X\times Y}$ is equal
  to $\tramePos{X}\times\tramePos{Y}$. It follows that $X\Par
  Y=(X\orth\otimes Y\orth)\orth$ is such that $(a,b)$ is positive in
  $X\Par Y$ iff $a$ or $b$ is positive and that $X\loli Y=X\orth \Par
  Y$ is such that $|X\loli Y|=|X|\times |Y|$ and
  $(a,b)\in\tramePos{X\loli Y}$ iff $a\in \tramePos{X}$ implies
  $b\in\tramePos{Y}$.
\item The \emph{of course} of $X$, $!X$ is the bipartite space of web
  $\Mfin{|X|}$ and of positive web
  $\tramePos{!X}=\Mfin{\tramePos{X}}$.  Thus a multiset $\mu$, element
  of $|!X|$, is negative iff $\mu$ contains at least one negative point
  of $X$. It follows that $?X=(!X\orth)\orth$ is such that an element
  $\mu$ of $|?X|$ is positive iff it contains at least one positive
  point of $X$.
\end{itemize}

\begin{rem}
  In contrast to the relational semantics, the bipartite semantics
  distinguishes $A$ and $A\orth$. In particular the multiplicative
  constants are distinct (this is not the case in hypercoherences).
\end{rem}

As usual a context $A_1,\ldots, A_n$ is interpreted by the same space
as the formula $A_1\Par\ldots\Par A_n$ . Interpretations of proofs are
defined as in the relational semantics. One easily verifies that a proof
is interpreted by a set of positive points (a clique). The categorical
structure of the bipartite semantics is derived from the one of the
relational semantics. Morphisms involved in natural transformations of the
semantics and morphisms obtained by functorial constructions, seen as
sets, are defined the same as in the relational semantics and it is
straightforward to verify that they actually contain only positive
points, so they are cliques. 

\subsection{Uniform bipartite semantics}

We introduce a uniform variant of the bipartite semantics as follows.
The uniform interpretation of exponentials is given by a comonad
$(\ue,\der[u],\dig[u])$ described below. The others categorical
constructions are the same as in the non uniform bipartite semantics. 

Setting $\Pos(X)=(\tramePos{X},\tramePos{X},\emptyset)$ for each
bipartite space $X$ and $\Pos(f)=f\cap(\tP{X}\times\tP{Y})$ for each
$f\in\Bip(X,Y)$ trivially makes $\Pos$ a functor. 

Since $f$ is a clique in $X\loli Y$, if $(a,b)\in f$ and $a\in \tP{X}$
then $b\in \tP{X}$. Hence $\Pos (f)$ is equal as a set with
$\Pos(\id[X])\comp f$, where $\Pos(\id[X])$ shall be seen by set
inclusion as a morphism from $\Pos(X)$ to $X$.  This can be used to
verify that $\Pos$ commutes with the composition: if $f\in\Bip(X,Y)$
and $g\in\Bip(Y,Z)$ then we have the following set equalities
\begin{align*}
\Pos(f)\comp\Pos(g)&=\Pos(f)\comp\id[\Pos({\id[Y]})]\comp g\\
&=\Pos(f)\comp g\\
&=\Pos(\id[X])\comp f\comp g\\
&=\Pos(f\comp
g).
\end{align*}

%% If $f\in\Bip(X,Y)$
%% and $g\in\Bip(Y,Z)$ then clearly $\Pos(f)\comp\Pos(g)\subseteq
%% \Pos(f\comp g)$ and, given iven an element $(a,c)$ of $\Pos(f\comp g)$
%% we use the fact that $f$ is a clique to deduce from $a\in\tramePos{X}$
%% that an element $b$ such that $(a,b)\in f$ and $(b,c)\in g$ is
%% positive which concludes to the equality $\Pos(f)\comp\Pos(g)=
%% \Pos(f\comp g)$.

The functor $\ue$ is defined by setting $\ue=\Pos !$ (where $!$ is the
functor ``of course'' of the non uniform bipartite semantics). Remark that
$\Pos !=!\Pos$. The natural transformations $\der[u]:\ue\dot{\to}\id$
and $\dig[u]:\ue\dot{\to} \ue\ue$ are defined by setting
$\der[u,X]=\{[a]\mid a \in \tP{X}\}$ (which is equal as a set to
$\der[\Pos(X)]$ and $\Pos\der[X]$) and $\dig[u]=\Pos(\dig)$ (one has
$\Pos!!=\ue\ue$). We have to verify that $\der[u]$ is actually a
natural transformation (for $\dig[u]$ this follows from the
definition). Let $f\in\Bip(X,Y)$. Then \[\der[u,X]\comp
f=\der[u,x]\comp \{(a,b)\mid a\in\tP{X}\}=\der[u,X]\comp
\Pos(\id[X])\comp f\] which is equal as a set with \[\Pos(\der[X]\comp
f)=\Pos(!f\comp \der[Y])=\ue f\comp\der[u,Y].\] And this concludes. The
followings commutative diagrams are the image by $\Pos$ of the
corresponding commutative diagrams for $(!,\der,\dig)$ in \Bip.
\[
\xymatrix{
{\ue E}\ar[dr]_{\ue I_{E}}\ar[r]^{\dig[ E]} 
& {\ue \ue E}\ar[d]^{\ue\der[u, E]}\\
&  {\ue E}
}\qquad
\xymatrix{
{\ue E} \ar[dr]_{\ue I_{E}}\ar[r]^{\dig[u,E]} 
& {\ue\ue E}\ar[d]^{\der[u,\ue E]}\\
   &  {\ue E}
}\qquad
\xymatrix{
  {\ue E}\ar[d]_{\dig[u,E]}\ar[r]^{\dig[u,E]} 
   & {\ue\ue E}\ar[d]^{\ue\dig[u,E]}\\
  {\ue\ue E}\ar[r]_{\dig[u,\ue E]} & {\ue\ue\ue E}
}
\]
Hence $(\ue,\der[u],\dig[u])$ is truly a comonad.

To achieve the verification that this setting form a new Seely
categorical semantics of linear logic one has to verify that the
adjunction induced by the comonad $(\ue,\der[u],\dig[u])$ is monoidal.
We won't check this in detail but it can be easily derived from the
fact that in \Rel the comonad $(!,\der,\dig)$ induces a monoidal
adjunction. Just remark that the isomorphisms $\ue(X\with Y)\cong
\ue(X)\otimes \ue(Y)$ and $\ue \top \cong 1$ hold and are natural
(since $\Pos(X')\otimes \Pos(Y')\cong\Pos(X'\otimes Y')$,
$\Pos(f)\otimes \Pos(g)$ is the same set as $\Pos(f\otimes g)$ and
$\Pos(1)=1$).

The interpretation of exponential rules in the uniform bipartite semantics
is obtained by a set restriction to the uniform web(s), as follows. 

The interpretations of the two rules contraction and
weakening, are, in fact, unchanged:
\[
\AXC{$\vdash\Gamma,?A,?A:f$}
\CONTR{\Gamma,?A: \{(\gamma,\mu_1+\mu_2)\mid (\gamma,\mu_1,\mu_2)\in f\}}
\DP\qquad
\AXC{$\vdash\Gamma:f$}
\AFFR{\Gamma,?A: \{(\gamma,[])\mid (\gamma)\in f\}}
\DP
\]
because in the sets produced by these two rules there are already only
negative points in $?A$. In the contraction rule, $\mu_1$ and $\mu_2$
contains only negative points of the interpretation of $A$, so does
$\mu_1+\mu_2$. And for the weakening rule we have that $[]$ is
negative in $?A$.

Thus, in contrast to the coherence space situation (for instance), the
construction associated with the contraction is the same as in the
relational semantics.

The interpretation of the dereliction rule is:
\[
\AXC{$\vdash\Gamma,A:f$}
\DERR{\Gamma,?A: \{(\gamma,[a])\mid (\gamma,a)\in f, a\in\tN{A}\}}
\DP
\]
(where $\tN{A}$ stands for $\tN{X}$ with $X$ interpreting $A$).  Remark
that this construction forgets some points relative to the non
uniform interpretation (contrarily to what happens in coherence spaces).

The interpretation of the promotion rule is:
\[
\AXC{$\vdash ?A_1,\ldots,?A_n,A:f$}
\PROMR{?A_1,\ldots,?A_n,!A:f^{\dag_u}}
\DP
\]
where $f^{\dag_u}$ is equal to the restriction to the uniform web of
the $f^\dag$ of the relational semantics. But again (contrarily to what
happens in coherence spaces) there is no need to restrict and so:
\[
f^{\dag_u}= f^\dag=\{(\sum_{j\in J} \mu_1^j,\ldots,\sum_{j\in J} \mu_n^j,
  [a_j\mid j\in J])\mid
[(\mu_1^j,\ldots,\mu_n^j, a_j)\mid j\in J]\in \Mfin(f)\}.
\]
This is because of two reasons. First, the sum of multisets of
negative points is a multiset of negative points, so $\sum_{j\in J}
\mu_i^j$ is in $\tN{?A_i}$, for each $i$. Second, since $f$ is a
clique, each element $(\mu_1,\ldots,\mu_n, a)$ of $f$ is
positive. Since each $\mu_i$ is negative in $?A_i$ it follows that $a$
is positive in $!A$ and so $[a_j\mid j\in J]\in\tP{!A}$.

\begin{rem}
  In the uniform bipartite semantics, each proof $\pi$ of a sequent
  $\vdash ?A$ is interpreted by the empty set. 
\end{rem}

To state this remark simply observe that the space interpeting $?A$
contains only negative points and that a clique is a set of positive
points.

\begin{exemple}
  The interpretation of the proof
  \[
  \AXR{\bot,1}
  \DERR{?1,\bot}
  \DP
  \]
  is the empty set. But the interpretation of the proof of
  Figure~\ref{fig:exsemunifpreuve} is the same as in the relational
  semantics.
\end{exemple}

Curiously enough the uniform bipartite semantics maps a lot of proofs to
the empty set. (But many other proofs are mapped on non trivial
subsets of their relational interpretations).

\subsection{Extensional collapses}

The basis types $\bool$ and $\nat$ are interpreted by the respective
bipartite spaces: 
\begin{align*}
  \bool&=(\{\true,\false\},\{\true,\false\},\emptyset) \et\\
  \nat&=(\bbbn,\bbbn,\emptyset).
\end{align*}

Simple types are interpreted by the same bipartite spaces in the
uniform bipartite semantics and in the non uniform bipartite semantics.
Moreover, the bipartite spaces interpreting simple types are purely
positive (every point of the web is positive) so every subset of the
web is a clique. Hence the uniform bipartite semantics, the non uniform
bipartite semantics and the relational semantics have the same extensional
collapse. We don't know any direct expression of this collapse.

\section{Non uniform \Kcoh semantics}
\label{sec:core}
\subsection{$K$-coherence spaces}
From now on, we shall assume that a subset $K$ of
$\bbbn\setminus\{0,1\}$ is given, and we call the corresponding
$\MultK$-coherence space a \Kcoh space.

% TODO décider quoi faire de ce paragraphe %%
% The notation $|X|$ for the web of $X$ might confuse the reader since
% we already used this notation for the relational interpretation of a
% formula. The confusion is deliberate since the web of the
% interpretation of a formula in the \Kcoh semantics will be its relational
% interpretation.

\subsection{Interpreting MALL... nothing new }

The interpretation of the multiplicative additive fragment of linear
logic (MALL) follows a standard pattern. 

% Linear negation is the exchange of coherence and incoherence:
% $X\orth=(|X|,\Rincoh,\Rcoh)$.

Both additive constants are the empty \Kcoh space: \[0=\top=(\emptyset,
\emptyset, \emptyset).\] Both multiplicative constants are the reflexive one
point \Kcoh space \[
1=\bot=(\{\ast\},\MultK(\{\ast\}), \MultK(\{\ast\})).
\]

Let $X_1$ and $X_2$ be two \Kcoh spaces.
  \begin{itemize}
  \item  The \Kcoh space $X_1\oplus X_2$ is
    defined by setting
    \begin{align*}
      |X_1\oplus X_2|&=|X_1| + |X_2|, \\
      \Rneutre[X_1\oplus X_2]&=\Rneutre[X_1]\uplus \Rneutre[X_2] \et\\
      \Rcohs[X_1\oplus X_2]&=\Rcohs[X_1]\uplus \Rcohs[X_2].
    \end{align*}
Of course
    $X_1\& X_2=(X_1\orth\oplus X_2\orth)\orth$. 
  
  \item The space $X_1\otimes X_2$ is defined as follows. We set
    $|X_1\otimes X_2|=|X_1|\times |X_2|$. For $i=1, 2$, let $\pi_i$
    be the canonical
    projections:
    \begin{align*}
      \pi_i:\MultK(|X_1\otimes X_2|)&\to\MultK(|X_i|)\\
      [(a^1_j,a^2_j)\mid j\in J]&\mapsto [a^i_j\mid j\in J].
    \end{align*}  
    Then for each $s\in\MultK(|X_1\otimes X_2|)$ we set
    \begin{align*}
      s\in \Rneutre[X_1\otimes X_2] &\ssi
      \pi_1(s)\in \Rneutre[X_1]\et\pi_2(s)\in\Rneutre[X_2]\\
      s\in\Rincohs[X_1\otimes X_2]&\ssi
      \pi_1(s)\in\Rincohs[X_1]{n}\ou\pi_2(s)\in\Rincohs[X_2]
    \end{align*}
    which suffices to determine $\Rneutre[X_1\otimes X_2]$,
    $\Rincohs[X_1\otimes X_2]$ and $\Rcohs[X_1\otimes X_2]$.  We also
    set ${X_1\Par X_2}=(X_1\orth\otimes X_2\orth)\orth$.
\end{itemize}

The \bindex{linear map} construction $\loli$ between \Kcoh spaces is
defined by setting $X\loli Y=X\orth\Par Y$. A \bindex{linear morphism}
from $X$ to $Y$, two \Kcoh spaces, is a clique of $X\loli Y$. Remark
that
\begin{align}
  s\in \Rcoh[X\loli Y]\ssi &\mbox{$\begin{cases}
\pi_1(s)\in\Rcoh[X]\implies
  \pi_2(s)\in\Rcoh[Y]\\ 
  \pi_1(s)\in\Rcohs[X]\implies \pi_2(s)\in\Rcohs[Y]
\end{cases}$}\label{loli:1}\\
\intertext{or equivalently, }
  s\in\Rcoh[X\loli Y] \ssi & \mbox{$\begin{cases}
      \pi_2(s)\in\Rincohs[Y]\implies \pi_1(s)\in\Rincohs[X] \\
      (\pi_1(s)\in\Rcoh[X]\et \pi_2(s)\in\Rneutre[Y])\implies
      \pi_1(s)\in\Rneutre[X].\end{cases}$}\label{loli:2}
\intertext{or equivalently,}
s\in \Rcoh[X\loli Y] \ssi
   & \mbox{$\begin{cases}
     \Rcoh[X]\pi_1(s)\implies \Rcoh[Y]\pi_2(s) \\ 
     \Rincoh[Y]\pi_2(s)\implies \Rincoh[X]\pi_1(s) 
     \end{cases}$}\label{loli:3}
\end{align}
We denote by \NCOHK the category whose objects are the \Kcoh spaces,
whose morphisms are the linear morphisms and where compositions and
identities are defined as in \Rel (one easily verifies that the
composition of a clique of $X\loli Y$ and a clique of $Y\loli Z$ is a
clique). For every $K'\subseteq K$, the corresponding categories come
naturally with forgetful functors $U_{K,K'}:\NCOHK\to\NCOHK[K']$ which
act as the identity on morphisms.
% and we also consider the
% cliques of a $K$-space ($\Cl_{K'}(X)=\Cl_K(U_{K,K'}X)$).
% Following (\ref{loli:1}) it comes that
% the relational composition $\comp$, defined by setting $f\comp
% g=\{(a,c)\mid(\exists b, (a,b)\in f\et (b,c)\in g\}$ for each
% $f\subseteq |A\loli B|$ and $g\subseteq|B\loli C|$ preserves the fact
% of being a linear morphism.  Moreover the identities of the relational
% semantics ($I_X=\{(a,a)\mid a\in |X|\}$) are linear morphisms thus the
% \Kcoh spaces and linear morphisms equipped with the relational
% composition and the relational identities form a category. We denote
% it \NCOHK. 

The \bindex{boolean type}, denoted by $\bool$ and represented by the formula
$1\oplus 1$ will be interpreted, in \NCOHK[\bbbn\setminus\{0,1\}], by the
uniform \Kcoh[\bbbn\setminus\{0,1\}] space whose web is $\{\true,\false\}$
and whose coherence is $\MultK[\bbbn\setminus\{0,1\}](\{\true\})\cup
\MultK[\bbbn\setminus\{0,1\}](\{\false\})$.

%TROIS CHOSES: valeurs de K: rel, coh et hc, foncteurs d'oublis et injections.

%le pb c'est que embed ce serait plutot pour (full) subcategory
%faithful = injective on each hom set
%full \forall a,b,f:Fa->Fb,\exists h:a->b Fh=f. surjectif su les hom-sets

\begin{proposition}[semantics of MALL]\label{prop:mall}
  For each $K\subseteq\bbbn\backslash\{0,1\}$, the category \NCOHK is a semantics
  of MALL. And for each $K'\subseteq K$ (in particular for $K'=\emptyset$) the
  functor $U_{K,K'}$ is logical w.r.t. the \NCOHK and \NCOHK[K'] MALL semantics
  (logicial means that, commutes to the interpretations of sequents and proofs).
%   of \Kcoh spaces and linear morphisms is
%   Cartesian and $\ast$-autonomous, \ie \NCOHK is a categorical semantics
%   of \mall, and the forgetful functor
%   $U:\NCOHK\rightarrow\Rel$ maps these structures to the
%   standard ones of \Rel.
\end{proposition}
The fact that \NCOHK is a semantics of MALL means that $\NCOHK$ is a
symmetric monoidal closed category (with $\otimes$ as tensor product
and $\loli$ as function space constructor) which is
$\ast$-auto\-nom\-ous ($\bot$ being the dualizing object), and
furthermore, has all finite products and coproducts (see
\cite{AmadioCurien} for precise definitions).  The proof, sketched
below, is a straightforward verification.

%DECOMPRESSE
 \begin{proof}
   The only thing to verify is that all the morphisms of \Rel which
   make \Rel a $\ast$-autonomous category, are cliques, that is
   linear morphism of the category \NCOHK. Provided $Y_1$ and $Y_2$ are
   of disjoint web, $Y_1\& Y_2$ equipped with the two projections
   \[p_i=\{(a,a)\mid a\in |Y_i|\}:(Y_1\& Y_2)\rightarrow Y_i\] for
   $i=1,2$ is the Cartesian product of $Y_1$ and $Y_2$ and if
   $f_1\in\NCOHK(X,Y_1)$ and $f_2\in\NCOHK(X,Y_2)$ then the
   \bindex{pairing} of $f_1$ and $f_2$ is just $(f_1\cup
   f_2)\in\NCOHK(X,Y_1\& Y_2)$. In fact one easily verifies that
   $p_1$, $p_2$ and $f_1\cup f_2$ are cliques. For each
   $f\in\NCOHK(X,Y)$, and $f'\in\NCOHK(X',Y')$, \[f\otimes
   f'=\{((a,a'),(b,b'))\mid (a,b)\in f, (a',b')\in f'\}\] is obviously
   a clique of $(X\otimes X')\loli (Y\otimes Y')$, thus the
   construction $\otimes$ on \Kcoh spaces is functorial. For each
   $X,Y,Z\in\NCOHK$, the isomorphisms $\unit : X\otimes 1\cong X$,
   $\ass : (X\otimes Y)\otimes Z\cong X\otimes (Y\otimes Z)$ and $\com
   : X\otimes X\rightarrow X\otimes X$ of \Rel given by
   \begin{align*}
   \unit&=\{((a,\ast),a)\mid a\in X\},\\
   \ass&=\{(((a,b),c),(a,(b,c)))\mid a\in X, b\in Y, c\in Z\}\text{ and}\\
   \com&=\{((a,b),(b,a))\mid a,b\in X\},
   \end{align*}
   are easily verified to be also isomorphisms in \NCOHK. So together
   with the functor $\otimes$ it gives a symmetric monoidal structure
   on \NCOHK which turns to be closed for the built-in object of
   morphisms $X\loli Y$.  Finally the dualising object $\bot$ is
   clearly such that $X\loli\bot\cong X\orth$ thus $X\orth\orth$ is
   obviously isomorphic to $X$.
 \end{proof}
%F_DEC
%   (of Cartesian product $\&$, of terminal object $\top$, of monoidal
%   product $\otimes$ with $1$ as neutral object and of dualizing object
%   $\bot$).

\begin{rem}[foliation]% consequence: 
% the forgetful functors $U_{K,K'}$ are logical.
  The coherence relation is \bindex[foliation]{foliated} with respects
  to the interpretation of MALL \ie for each formula $A$ of MALL the
  coherence relations on multisets of cardinality $n$ in the
  interpretation of $A$ is totally determined by the coherence
  relations on multisets of cardinality $n$ in the interpretation of
  the sub-formulae of $A$.  In fact, this is exactly by constructing
  independently each coherence relation of level $k$ for $k\in K$ in
  the Bucciarelli-Ehrhard machinery that the \Kcoh spaces semantics has
  been obtained, so this remark also holds for the linear logic semantics
  with the exponential provided by this machinery.  Anticipating a
  bit, it will also hold for the new exponential construction we
  present (and the forgetful functors $U_{K,K'}$ will still be logical
  in LL).
\end{rem}

\subsection{Exponentials}
Using the constructions presented by Bucciarelli and Ehrhard
in~\cite{phaseexp}, one can define exponentials for \Kcoh
spaces. 

This gives a semantics which accepts a variant
%% \[
%% \begin{array}{r l l l l}
%%   f=\{ &(([],&[\true,\true],&[\false,\false]),&\true), \\
%%          &(([\false,\false],&[],&[\true,\true]),&\true), \\
%%          & (([\true,\true],&[\false,\false],&[]),&\true)\}
%% \end{array}
%% \]
of the well known Berry's example of a stable and non sequential
function from $\bool\times\bool\times\bool$ to $\bool$. Of course
both the standard (set based) hypercoherence semantics and the
multiset based hypercoherence semantics we use here reject such first
order non-sequential functions.

%DECOMPRESSE
The corresponding ``of course'' operation is denoted by $\E$ and
defined as follows. We set $|\E X|=\Mfin(|X|)$. A multiset $[x_i\mid
1\leq i\leq k]\in\MultK(|\E X|)$ (so that $k\in K$) is strictly
incoherent in $\E X$ iff there exists a multiset $[a_j\mid 1\leq j\leq
k]\in\MultK(|X|)$ which is strictly incoherent in $X$ and satisfies
\[{[a_j\mid 1\leq j\leq k]\leq\sum_{i=1}^k x_i}.\] If such a multiset
$[a_j\mid 1\leq j\leq k]$ does not exist, $[x_i\mid 1\leq i\leq k]$ is
coherent and then, it is strictly coherent exactly when $\sum_{i=1}^k
x_i$ is \emph{star-shaped} that is when there exists
$a\in|\sum_{i=1}^k x_i|$ such that
 \begin{equation*}    
    \forall (a_j)_{1\leq j\leq k} \in
    |X|^k, ([a_j\mid 1\leq j\leq k]\leq\sum_{i\in I} x_i \et
    a_k=a)\implies [a_j\mid 1\leq j\leq k]\in\Rcohs\,.
 \end{equation*}

 For instance $\E\bool$ is given by:
 \begin{align*}
   [x_i\mid i\in I]\in\Rincohs[\E\bool] &\ssi \sum_{i\in I} x_i =
   p[\true]+q[\false]
   \text{ with } p,q>0 \et p+q\geq \card I \\
   [x_i\mid i\in I]\in\Rcohs[\E\bool] &\ssi \sum_{i\in I} x_i =
   p[\true]+q[\false] \text{ with } 1\leq p+q< \card I \\
   \text{hence } [x_i\mid i\in I]\in\Rneutre[\E\bool] & \ssi\sum_{i\in I}
   x_i = [], k[\true] \ou k[\false] \text{ with }k\geq \card I \\
 \end{align*}
 Consider the following subset of
 $|(\E\bool\otimes\E\bool\otimes\E\bool)\loli\bool|$:
 \[
 \begin{array}{r l l l l}
   f=\{ &(([],&[\true,\true],&[\false,\false]),&\true), \\
          &(([\false,\false],&[],&[\true,\true]),&\true), \\
          & (([\true,\true],&[\false,\false],&[]),&\true)\quad\}.
 \end{array}
 \]
 It is a variant of the well known Berry's example of a stable and non
 sequential function from $\bool\times\bool\times\bool$ to $\bool$.
 This function is not a morphism in the multiset based hypercoherence
 semantics but the \Kcoh[\bbbn\backslash\{0,1\}] semantics with the
 $\E$ exponential accepts it \footnote{Remark that the same function
   with $[\true]$ instead of $[\true,\true]$ and $[\false]$ instead of
   $[\false,\false]$, is rejected by this semantics.}. Indeed, for
 each multiset $s\in\MultK(f)$, on has $\pi_2(s)\in\MultK(\{\true\})$
 thus $\pi_2(s)\in \Rneutre[\bool]$.  With respect to
 Equation~\ref{loli:1}, the only way for $s$ to be strictly
 incoherent in $(\E\bool\otimes\E\bool\otimes\E\bool)\loli\bool$ is to
 have $\pi_1(s)\in\Rcohs[]$ in
 $\E\bool\otimes\E\bool\otimes\E\bool$. But if $m_1$, $m_2$ and $m_3$
 are the respective numbers of occurrences of points of $f$ in $s$
 then
 \begin{itemize}
 \item if only one of the $m_i$ is non-zero then each of the projection
   of $\pi_1(s)$ on the three arguments is neutral so $\pi_1(s)$ is
   neutral;
 \item if exactly two of the $m_i$ are non-zero (say $m_1$ and $m_2$)
   then the empty multiset does not occurs in one of the three
   projections of $\pi_1(s)$ (here the third) thus the sum of the
   multisets occurring in this projection contains enough $\true$ and
   $\false$ (here $2m_1\false$ and $2m_2\true$) comparatively to its
   cardinality (here $m_1+m_2$) as to make it strictly incoherent in
   $\bool$. Hence $\pi_1(s)$ is surely strictly incoherent;
  
 \item finally, if none of the $m_i$ is zero then $\pi_1(s)$ is coherent
   iff each of its three projections on $\E\bool$ is coherent. That
   is iff
 \begin{align*}
   m_1+m_2+m_3 &> 2 m_2+ 2 m_3
   \text{ (coherence on the first argument)}\\
   m_1+m_2+m_3 &> 2 m_1+ 2 m_3
   \text{ (coherence on the second argument)}\\
   m_1+m_2+m_3 &> 2 m_1+ 2 m_2 \text{ (coherence on the third
     argument)}
   \end{align*}
   but we will then have that $3(m_1+m_2+m_3)>4(m_1+m_2+m_3)$ which is
   impossible.
 \end{itemize}
 Thus $f$ is definitely a clique of
 $(\E\bool\otimes\E\bool\otimes\E\bool)\loli\bool$.
% contrarily to the hypercoherence situation.
%F_DEC

 The really surprising fact is that one can easily try to correct this
 by choosing another definition for the coherence relations of the
 exponential construction and obtain in that way a new semantics of linear
 logic.  Among these variants for the exponentials there is a
 \emph{most general one} in a sense which will be made precise in
 Theorem~\ref{theo:free} and Corollary~\ref{cor:max}. Indeed the
 definition is guided by the need of Theorem~\ref{theo:free}.

First of all we adapt the notion of section of hypercoherences to the
\Kcoh spaces setting. 

\begin{definition}
  If $\mu=[x_i\mid i\in I]$ is a multiset of finite sets or of
  multisets and if $s$ is another multiset we say that $s$ is a
  \bindex{section} of $\mu$ and we write $s\msect \mu$ when there
  exists a family $(a_i)_{i\in I}$ such that $\forall i\in I, a_i\in
  x_i$ and $s=[a_i\mid i\in I]$ (in particular $s$ and $\mu$ have the
  same cardinality).
\end{definition}

The notion of section between \emph{sets} we used until now for
hypercoherences can be rephrased by saying that a set $s$ is a
section of a set $x$ iff there exists two multisets $\mu$ and $\nu$
such that $\nu \msect \mu$, $\supp{\mu}=x$ and $\supp{\nu}=s$. We use
the same name (section) but a different notation for the two notions:
$\sect$ between sets, $\msect$ between multisets.

\begin{definition}\label{def:exp}
  For each \Kcoh space $X$ we define the \Kcoh space $!X$ as follows.
  Its web is $|!X|=\Mfin(|X|)$ and for each element $[x_i\mid i\in I]$ 
  of $\MultK(|!X|)$ we set:
\begin{align}    
  [x_i\mid i\in I]\in\Rincohs[!X]&\ssi\exists (a_i)_{i\in I}, [a_i\mid i\in I]\in\Rincohs[X]\et \forall i\in I, a_i\in x_i\label{exp:incoh}\\
  \intertext{ and } [x_i\mid i\in I]\in\Rneutre[!X]
  &\ssi \begin{cases} [x_i\mid i\in I]\notin\Rincohs[!X] \et \\
    \exists (a^j_i)^{j\in J}_{i\in I}, \mbox{ $\begin{cases}
        \forall i\in I,[a^j_i\mid j\in J]= x_i \et\\
        \forall j\in J,[a^j_i\mid i\in I]\in\Rneutre[X]
  \end{cases}$
}
\end{cases}\label{exp:neutre}
\end{align}
We also define $?X$ by setting $?X=(!X\orth)\orth$.
\end{definition}

When $K=\emptyset$, the exponential construction on objects is the
standard exponential of \Rel.

\begin{exemple}
\label{ex:berry}
The coherence in $!\bool$ is as follows. For $\mu\in\MultK(|\bool|)$
\begin{align*}
  \mu\in\Rcoh[!\bool]&\ssi 
\begin{cases}
  \supp \mu \subset \Mfin(\{\true\}) \ou \\
  \supp \mu \subset \Mfin(\{\false\}) \ou \\
  []\in\mu
\end{cases}\\
\mu \in \Rneutre[!\bool]&\ssi \supp \mu = \{k [\true]\} \ou \{k
[\false]\}, k\in\bbbn.
\end{align*}

Thanks to this exponential, our variant of the Berry's example is
successfully rejected : if $3\in K$, then $f$ (as previously defined)
is not a clique.  Take $[a, b, c]\in\MultK(f)$ where $a$, $b$ and $c$
are the three points of $f$, then each of the three projections of
$[a, b, c]$ on $!\bool$ is strictly coherent so is the projection on
$!\bool\otimes !\bool\otimes !\bool$ but the projection on $\bool$ is
neutral thus $[a, b, c]\in \Rincohs$.
\end{exemple}

\begin{exemple}\label{ex:G}
  Consider the \Kcoh[K] space $G$ with web $|G|=\{a,b,c\}$ and such
  that if $u\in\Mfin(|G|)$ then: $u\in\Rneutre[G]$ iff $\supp{u}$ is a
  singleton, $u\in\Rcohs[G]$ iff $\card\supp{u}=2$ and
  $u\in\Rincohs[G]$ iff $\supp{u}=\{a,b,c\}$.  The space $G$ is in
  fact the sub-space of $\bool^3\to\bool$ of web (the variant of) the
  Berry's example $f$ above.
  
  Suppose $2\in K$. All the sections of $[[a],[b,c]]$ are coherent in
  $G$ moreover $[a]$ and $[b,c]$ have not the same cardinality. So
  $[[a],[b,c]]\in\Rcohs[!G]$. Now suppose $3\in K$. Then
  $[[a],[b,c],[b,c]]$ admits the strictly incoherent section $[a,b,c]$
  but $[[a],[a],[b,c]]$ not and so $[[a],[b,c],[b,c]]\in\Rincohs[!G]$
  but $[[a],[a],[b,c]]\in\Rcohs[!G]$. So the coherence relations of
  $!G$ depends on multiplicities.  
  
  For each $k\in K$ such that $k\geq 3$, each $m\in\MultK[\{k\}](|G|)$
  such that $\supp{m}=\{[a,b],[a,c]\}$ is strictly incoherent in $!G$
  but if $2\in K$, $[[a,b],[a,c]]\in\Rcohs[!G]$ (all the sections of
  $[[a,b],[a,c]]$ are coherent in $G$ and $b$ is not neutral with any
  element of $[a,c]$).
 
  Finally $[[a,b,c],[a,b,c],[a,b,c]]$ is an example of a non neutral
  (strictly incoherent, here) multiset in $!G$ of support a singleton.
\end{exemple}

\begin{proposition}[semantics of LL]\label{prop:ll}
  Any category \NCOHK with the exponentials of Definition~\ref{def:exp} is a semantics of linear logic (see \cite{AmadioCurien}
  and \cite{whatillcatmod}) and for each $K'\subseteq K$ (in
  particular for $K'=\emptyset$) the functor $U_{K,K'}$ is logical
  w.r.t. the \NCOHK and \NCOHK[K'] LL semantics.
\end{proposition}

\begin{proof}  
%%   For the functoriality of $!$ as for its comonad structure we
%%   just follow the standard \Rel construction. That is, for each
%%   morphism $f\in\NCOHK(X,Y)$ we set:
%%   \[
%%   !f=\{([a_1,\ldots,a_n],[b_1,\ldots,b_n])\mid n\in\bbbn, \forall
%%   i\in\{1,\ldots, n\}, (a_i,b_i)\in f\}
%%   \]
%% and for each \Kcoh space $X$ we set:
%% \begin{align*}
%%   \der[X]&=\{([a],a)\mid a\in |X|\} \;\et\\ \dig[X]&= \{(\sum_{1\leq
%%     i\leq n} x_i,[x_1,\ldots,x_n])\mid n\in\bbbn,[x_1,\ldots, x_n]\in
%%   |!!X|\}.
%% \end{align*}  
  We equip \NCOHK with the comonad structure $(!,\der[],\dig[])$ of
  \Rel. We exploit the fact that the required commutative diagrams
  already hold in \Rel and therefore also in \NCOHK. Hence to check
  that $(!,\der[],\dig[])$ is really a comonad we only need to prove
  that if $f$ is a clique of $X\loli Y$ then $!f$ is a clique of
  $!X\loli !Y$, that $\der[X]$ is a clique of $!X\loli X$ and that
  $\dig[X]$ is a clique of $!X\loli !!X$. The same for the monoidality
  of the adjunction: we only need to ckeck that the \Rel isomorphisms
  $!\top\cong 1$ and $!(X\& Y)\cong{!}X\otimes !Y$ are cliques (in
  both directions).

%%   To show that $(!,\der[],\dig[])$ is a comonad of \NCOHK, exploiting
%%   that all the required equalities (commutative diagrams) already hold
%%   in \NCOHK[\emptyset], and therefore also in \NCOHK, we only need to
%%   prove that $!f$ is a clique of $!X\loli !Y$, that $\der[X]$ is a
%%   clique of $!X\loli X$ and that $\dig[X]$ is a clique of $!X\loli
%%   !!X$.
  
  Let $[(x_j,y_j)\mid j\in J]\in\MultK(!f)$. If $[b_j\mid j\in
  J]\msect[y_j\mid j\in J]$ then by construction of $!f$ there exists
  $[a_j\mid j\in J]$ such that ${[(a_j,b_j)\mid j\in J]\in\MultK(f)}$
  and $[a_j\mid j\in J]\msect[x_j\mid j\in J]$. Remark that since $f$
  is a clique, we have ${[(a_j,b_j)\mid j\in J]\in\Rcoh[X\loli Y]}$. In
  particular, if $[b_j\mid j\in J]\in\Rincohs[Y]$ then ${[a_j\mid j\in
  J]\in\Rincohs[X]}$.  Hence if $[y_j\mid j\in J]$ admits a strict
  incoherent section then $(x_j)_{j\in J}$ admits one too. So
  \[[y_j\mid j\in J]\in \Rincohs[!Y]\implies [x_j\mid j\in J]\in
  \Rincohs[!X].\] Now suppose $[x_j\mid j\in J]\in\Rcoh[!X]$ and
  $[y_j\mid j\in J]\in\Rneutre[!Y]$.  We must prove that $[x_j\mid
  j\in J]\in\Rneutre[!X]$. There exists $(b^i_j)_{(i,j)\in I\times J}$
  such that \[\forall j\in J, y_j=[b^i_j\mid i\in I]\text{ and
  }\forall i\in I, [b^i_j\mid j\in J]\in \Rneutre[Y].\] By
  construction of $!f$ there exists $(a^i_j)_{(i,j)\in I\times J}$
  such that \[\forall (i,j)\in I\times J, (a^i_j,b^i_j)\in f\text{ and }
  \forall j\in J, x_j=[a^i_j\mid i\in I].\]  Since $[x_j\mid j\in
  J]\in\Rcoh[!X]$ for each $i\in I$, $[a^i_j\mid j\in J]\in\Rcoh[X]$.
  But, for each $i\in I$, $[(a^i_j,b^i_j)\mid j\in
  J]\in\MultK(f)\subseteq \Rcoh[X\loli Y]$ and $[b^i_j\mid j\in
  J]\in\Rneutre[Y]$ so $[a^i_j\mid j\in J]\in\Rneutre[X]$, for each
  $i\in I$.  Finally $[x_j\mid j\in J]\in\Rneutre[!X]$ which concludes
  the proof that $!f$ is a clique.

  The fact that $\der[X]$ is a clique is straightforward. We now prove
  that $\dig[X]$ is a clique of $!X\loli !!X$. Let $[(\sum_{i\in I_j}
  x^j_i,[x^j_i\mid i\in I_j])\mid j\in J]\in\MultK(\dig[X])$. 
  
  Suppose $[[x^j_i\mid i\in I_j]\mid j\in J]\in \Rincohs[!!X]$. Then
  this multiset admits a section $[y_j\mid j\in J]$ strictly
  incoherent in $!X$.  Hence this section $[y_j\mid j\in J]$ admits a
  section $[a_j\mid j\in J]$ strictly incoherent in $X$. Clearly this
  last section is also a section of $[\sum_{i\in I} x^j_i \mid j\in
  J]$ so this multiset is strictly incoherent in $!X$.
  
  Now suppose $[\sum_{i\in I_j} x^j_i\mid j\in J]\in\Rcoh[!X]$ and
  $[[x^j_i\mid i\in I_j]\mid j\in J]\in \Rneutre[!!X]$.  Then there
  exists a family $(y^j_i)^{j\in J}_{i\in I}$ such that: for all $j\in
  J$, $[y^j_i\mid i\in I]$ equals $[x^j_i\mid i\in I_j]$ (so $I=I_j$ and
  $\sum_{i\in I} y^j_i=\sum_{i\in I_j} x^j_i$); and for all $i\in I$,
  $[y^j_i\mid j\in J]\in\Rneutre[!X]$. Hence for each $i\in I$, there
  exists a family $(a^j_{i,l})^{j\in J}_{l\in L_i}$ such that for all
  $j\in J$, $y^j_i=[a^j_{i,l}\mid l\in L_i]$ and such that for all
  $l\in L_i$, $[a^j_{i,l}\mid j\in J]\in\Rneutre$. Without any lost of
  generalities the $L_i$ can be chosen pairwise disjoint. Setting
  $L=\cup_{i\in I} L_i$, we then have $\sum_{l\in L}
  a^j_{l}=\sum_{i\in I_j} x^j_i$ and for all $l\in L$, $[a^j_{l}\mid
  j\in J]\in\Rneutre$.  Hence $[\sum_{i\in I_j} x^j_i\mid j\in
  J]\in\Rneutre[!X]$.

%\fbox{illustration}
%%% TODO version longue : une image. 
  
  The set $\{([],\ast)\}$ is a clique of $!\top\loli 1$ and the set
  $\{(\ast,[])\}$ is a clique of $1\loli{!}\top$ so $!\top\cong 1$.
  We now prove that $!(X\& Y)\cong{!}X\otimes !Y$, for each $X$ and
  $Y$.  The graph $f$ of the bijection map
\[\left\{
\begin{array}{r c l}
      \Mfin(|X|)\times\Mfin(|Y|) & \to & \Mfin(|X\& Y|)\\
      (x, y)& \mapsto & x+y 
\end{array}\right.
\]
is a relational isomorphism. It remains to prove that $f$ is a clique
of $!(X\& Y)\loli (!X\otimes !Y)$ and that its transpose is
a clique of $(!X\otimes !Y)\loli!(X\& Y)$.  Consider a
multiset $[((x_i,y_i),x_i+y_i)\mid i\in I]\in \MultK(f)$. Since an
element of $\Rincohs[X\& Y]$ is either an element of $\Rincohs[X]$ or
an element of $\Rincohs[Y]$, a section $s$ of $[x_i+y_i\mid i\in I]$
is strictly incoherent in $X\& Y$ iff $s$ is a strictly incoherent
section of $[x_i\mid i\in I]$ or of $[y_i\mid i\in I]$. It follows
that
\[
[x_i+y_i\mid i\in I]\in\Rincohs[!(X\& Y)]\iff [(x_i,y_i)\mid i\in
I]\in\Rincohs[!X\otimes !Y].
\]
An element of $\Rneutre[X\& Y]$ is either an element of $\Rneutre[X]$
or an element $\Rneutre[Y]$. Hence, if $[x_i+y_i\mid i\in I]$ is
neutral in $!(X\& Y)$, there exists a family $(c^j_i)^{j\in J}_{i\in
  I}$ such that for each $j\in J$, $[c^j_i\mid i\in
I]\in\Rneutre[X\with Y]$ and such that $J=J_X+J_Y$ with, for each
$i\in I$, $[c^j_i\mid j\in J_X]=x_i$ and $[c^j_i\mid j\in J_Y]=y_i$
and this family splits into two families, the first one corresponding to
the neutrality of $[x_i\mid i\in I]$ in $!X$ and the other one to the
neutrality of $[y_i\mid i\in I]$ in $!Y$.  Consequently the neutrality
of $[x_i+y_i\mid i\in I]$ in $!(X\& Y)$ implies the neutrality of
$[(x_i,y_i)\mid i\in I]$ in $!X\otimes !Y$.  The converse is
straightforward. So the required isomorphisms $!\top\cong 1$ and
$!(X\with Y)\cong !X\otimes !Y$ holds. At last we obtain for free that
this two isomorphisms are naturals and that the adjonction involved by
the comonad is monoidal (see \cite{whatillcatmod}) just by using the
fact that this is already the case in \Rel.
\end{proof}

\subsection{The \emph{of course} is the co-free commutative $\otimes$-comonoid}
\label{sec:free}

A commutative comonoid on a symmetric monoidal category $\C$, with
respect to a monoidal structure $(\otimes,\sym[],\ass[],\unit[])$, is
a $3$-tuple $M=(\UL{M},u_M,\mu_M)$, where
$\UL{M}\in\C$, 
$u_M\in\C(\UL{M},1)$
and $\mu_M\in \C(\UL{M},\UL{M}\otimes\UL{M})$,
% (associativity)
% $\id[\UL{M}]\otimes\mu_M=(\mu_M\otimes \id[\UL{M}])\comp
% \ass[\UL{M},\UL{M},\UL{M}]$; (neutrality)
% $\mu_M\comp(\id[\UL{M}]\otimes u_M)=\unit[\UL{M}]$ and (commutativity)
% $\mu_M\comp\sym[\UL{M}]=\mu_M$. 
such that the following diagrams
commute:
\begin{gather*}
\shortstack{
\xymatrix{
{\UL M\otimes \UL M} \ar[drr]_{\mu_M\otimes\id[\UL M]}\ar[rr]^{\id[\UL
  M]\otimes \mu_M} 
 & & {\UL M\otimes(\UL M\otimes \UL M)}\ar[d]^{\ass[\UL M, \UL M, \UL M]}\\
 & &  {(\UL M\otimes\UL M)\otimes \UL M}
}\\
associativity
}
\quad
\shortstack{
\xymatrix{
{\UL M} \ar[dr]_{\unit[\UL M]}\ar[r]^{\mu_M} 
& {\UL M\otimes \UL M}\ar[d]^{\id[\UL M]\otimes u_M}\\
   &  {\UL M\otimes 1}
}\\ 
neutrality}\quad
\shortstack{
\xymatrix{
{\UL M} \ar[dr]_{\mu_M}\ar[r]^{\mu_M} 
& {\UL M\otimes \UL M}\ar[d]^{\sym[\UL
  M\otimes \UL M]}\\
&  {\UL M\otimes \UL M}
}\\
commutativity
}
\end{gather*}
A comonoid morphism $f$ from
$(\UL M,u_M,\mu_M)$ to $(\UL N,u_N,\mu_N)$ is a morphism $f\in\C(M,
N)$ such that the following diagrams commute:
% commuting with the co-units and the co-multiplications
% that $f\comp u_N=u_M$ and $f\comp\mu_N=\mu_M\comp (f\otimes f)$.
\begin{gather*}
\xymatrix{
\UL M\ar[dr]_{u_M}\ar[r]^f & \UL N\ar[d]^{u_N}\\
&1
}\qquad
\xymatrix{
\UL M\ar[d]_{\mu_M}\ar[r]^f & \UL N\ar[d]^{\mu_N}\\
\UL M\otimes \UL M\ar[r]^{f\otimes f} & \UL N\otimes\UL N
}
\end{gather*}

In each categorical semantics $\C$ of linear logic the ``of course''
naturally provides a commutative comonoid $(!X,\aff,\cont)$ for each
object $X$: $\aff[X]$ is $!\top_X$ where $\top_X$ is the unique morphism
of $\C(X,\top)$ and $\cont[X]$ is $(!\langle \id[X], \id[X]\rangle)\comp
e_X$ where $\langle \id[X], \id[X]\rangle$ denotes the pairing of the
identity with itself and where $e_X$ is the isomorphism $!(X\with
X)\cong !X\otimes !X$. Moreover for
each $f\in\C(X,Y)$, $!f$ is a $\otimes$-comonoid morphism between
$(!X,\aff[X],\cont[X])$ and $(!Y,\aff[Y],\cont[Y])$.

In \NCOHK, $\aff[X]=\{([],\ast)\}$ and
$\cont[X]=\{(x_1+x_2,(x_1,x_2))\mid x_1,x_2\in |!X|\}$.

%$\cont:!X\rightarrow !X\otimes !X$,
%$\aff :!X\rightarrow 1$

A commutative comonoid $(F,u_F,\mu_F)$ is said to be co-free over an
object $X$ of $\C$ when there exists a morphism $d\in\C(F,X)$ such
that for each commutative comonoid $(A,u_A,\mu_A)$, and for each
$f\in\C(A,X)$ there exists a unique comonoid morphism $f_{\ast}$ from
$(A,u_A,\mu_A)$ to $(F,u_F,\mu_F)$ such that ${f_{\ast}\comp d}=f$.
\begin{gather*}
\xymatrix{
(A,u_A,\mu_A)\ar[dr]_f \ar[r]^{f_\ast}& (F,u_F,\mu_F)\ar[d]^d\\
& X}
\end{gather*}
By extension the ``of course'' $!$ is said to be the co-free commutative
$\otimes$-comonoid or, for short, to be co-free, when for each
commutative comonoid $(A,u_A,\mu_A)$, for each $X\in\C$ and for each
$f\in\C(A,X)$ there exists a unique comonoid morphism
$f_{\ast}:(A,u_A,\mu_A)\to (!X,\aff[X],\cont[X])$ such that
\[{f_{\ast}\comp\der[X]}=f.\] 

\begin{rem}
  If $!$ is co-free then $f_{\ast}=\Id_{\ast}\comp !f$ where $\Id$ is
  the identity morphism in $\C(A,A)$.
\end{rem}

\begin{lemme}\label{lem:Iast}
  In \Rel the exponential is co-free. Moreover if $(A,u_A,\mu_A)$ is a
  commutative $\otimes$-comonoid in $\Rel$ then $(a,x)\in
  (\Id_{A})_{\ast}$ iff if $(a_i)_{1\leq i\leq n}$ is such that
  $[a_1,\ldots,a_n]=x$ then $\exists (b_i)_{0\leq i\leq n}$ such that
  $b_0=a$, $(b_i,(a_{i+1},b_{i+1}))\in\mu_A$ for each $i<n$, and
  $(b_n,\ast)\in u_A$.
\end{lemme}

\begin{theo}[co-free]\label{theo:free}
  The ``of course'' $!$ is the co-free commutative $\otimes$-co\-mon\-oid of
  \NCOHK and the forgetful functor
  $\U_{K,\emptyset}:\NCOHK\rightarrow\Rel$ maps this structure to the
  standard one.
\end{theo}

\begin{proof}
  We prove that for each commutative comonoid $(A,u_A,\mu_A)$ of
  \NCOHK for each $X\in\NCOHK$ and for each $f\in\NCOHK(A,X)$, there
  exists a unique comonoid morphism $f_{\ast}:(A,u_A,\mu_A)\rightarrow
  (!X,\der,\cont)$ such that $f_{\ast}\comp\der=f$.
  
  But if there is such an $f_{\ast}$ in \NCOHK then $\UK[f_{\ast}]$ is
  a comonoid morphism from $(\UK[A],\UK[u_A],\UK[\mu_A])$ to
  $(\UK[!X],\UK[{\der[X]}],\UK[\cont])$ and
  \[
\UK[f_{\ast}]\comp\UK[{\der[X]}]=\UK[f].\]  As
  $(\UK[!X],\UK[{\der[X]}],\UK[\cont])$ is the co-free
  $\otimes$-comonoid in \Rel this means that, in \Rel,   \begin{align*}
  \UK[f_{\ast}]&=\UK[f]_{\ast}.\\
\intertext{Moreover}
    \UK[f]_{\ast}&=\UK[\Id]_{\ast}\comp !\UK[f]\\
\intertext{and}
    !\UK[f]&=\UK[!f]. 
  \end{align*}
  So the only thing to prove is that $\UK[\Id]_{\ast}$ is a clique of
  $\Cl(A\loli !A)$.
  
  Let $[(a^i,[a^i_1,\ldots,a^i_{n_i}])\mid i\in I]$ be an element of
  $\MultK(\Id_{\ast})$.  Then, using Lemma~\ref{lem:Iast}, for each
  $i\in I$, let $(b^i_j)_{0\leq j\leq n_i}$ be a family such that
  $b^i_0=a^i$, $(b^i_j,(a^i_{j+1},b^i_{j+1}))\in\mu_A$ for each
  $j<n_i$, and $(b^i_{n_i},\ast)\in u_A$.
  
  Suppose $[[a^i_1,\ldots,a^i_{n_i}]\mid i\in I]\in\Rincohs[!A]$ then
  this multiset admits a strict incoherent section. Up to a choice of
  an adequate indexation of the multiset $[a^i_1,\ldots,a^i_{n_i}]$,
  we can suppose without any loss of generality that this section is
  $[a^i_1 \mid i\in I]$.  Remark that due to the existence of a
  section, none of the $n_i$ is zero.  We then have
  $[(a^i,(a^i_1,b^i_1))\mid i\in I]\in\MultK(\mu_A)$ with $[a^i_1 \mid
  i\in I]\in \Rincohs[A]$. Hence $[(a^i_1,b^i_1)\mid i\in
  I]\in\Rincohs[A\otimes A]$. And, since $[(a^i,(a^i_1,b^i_1))\mid i\in
  I]$ must be coherent for $\mu_A$ to be a clique of $A\loli(A\otimes
  A)$, we then have $[a^i \mid i\in I]\in\Rincohs[A]$.
  
  Now suppose $[a^i\mid i\in I]\in\Rcoh[A]$ and
  $[[a^i_1,\ldots,a^i_{n_i}]\mid i\in I]\in\Rneutre[A]$.  According to
  the definition of neutrality in the ``of course'', all the $n_i$ are
  equal, say $n_i=n (\forall i\in I)$, and, up to an appropriate
  re-indexing, $[a^i_j\mid i\in I]\in\Rneutre[A]$, for each $1\leq
  j\leq n$.  Since $[(b^i_n,\ast) \mid i\in
  I]\in\MultK(u_A)\subseteq\Rcoh[A\loli 1]$ and $[\ast\mid i\in
  I]\in\Rneutre[1]$, this means that $[b^i_n\mid i\in
  I]\in\Rincoh[A]$.  Now suppose $[b^i_{k+1}\mid i\in I]\in\Rincoh[A]$
  for a certain $k<n$, then using $[a^i_{k+1}\mid i\in
  I]\in\Rneutre[A]$ and $[(b^i_k,(a^i_{k+1},b^i_{k+1}))\mid i\in
  I]\in\MultK(\mu_A)$ it follows that $[b^i_{k}\mid i\in
  I]\in\Rincoh[A]$. Thus, for all $j\leq n$, $[b^i_{j}\mid i\in
  I]\in\Rincoh[A]$ and in particular $[b^i_{0}\mid i\in I]=[a^i\mid
  i\in I]$ is then proved to be both coherent and incoherent, that is to
  be neutral. So $\Id_{\ast}$ is a clique.
\end{proof}

Consider a sub-category $\C$ of \NCOHK which is a categorical semantics of
intuitionistic multiplicative exponential linear logic\footnote{We do
  not require $\C$ satisfies more, but a typical $\C$ for our purpose
  will be a new Seely category where the multiplicative additive and
  orthogonal constructions are the ones of \NCOHK and so one should
  have a semantics for the full linear logic fragment, where the
  exponentials are given by a comonad.}.  Let $E$ be the operation
modeling the ``of course'' on objects in $\C$. We shall say that this
semantics is \emph{multiset based} if for each $X\in\C$:

\begin{itemize}
\item the web of $E(X)$ is made of multisets of points of the web of
  $X$ (\ie $|E(X)|\subseteq\Mfin(|X|)$);
\item the commutative comonoid structure provided with $E(X)$ by the
  semantics is defined by $
  \aff[X]'= \{([],\ast)\}$ (of type $E(X)\to 1$) and
  \[
  \cont[X]'=\{(x_1+x_2,(x_1,x_2))\mid x_1+x_2\in|E(X)|\et
  x_1,x_2\in\Mfin(|X|)\}\] (of type $E(X)\to E(X)\otimes E(X)$);

\item the associated dereliction morphism is \[{\der[X]'=\{([a],a)\mid
  a\in |X|\}} \text{ (of type $E(X)\to X$)}.\]
\end{itemize}

\begin{corollaire}[maximality of the co-free ``of course'']\label{cor:max}
  If a sub-monoidal category $\C$ of $\NCOHK$ is a multiset based LL
  semantics, of ``of course'' $E$ then, for each object $X\in\C$,
  \begin{gather}
    \Rcoh[E(X)]\subseteq \Rcoh[!X] \label{cormax:1}\\ \et\notag \\
    \Rcohs[E(X)]\subseteq \Rcohs[!X].\label{cormax:2}
\end{gather}
\end{corollaire}
``Sub-monoidal category'' means that $\C$ is a sub-category of \NCOHK
equipped with the same symmetric monoidal structure as \NCOHK. 
% In fact to be a categorical semantics of \LiLo, $\C$ must provides $E$
% with a comonad structure. Thus for each $X\in\C$ there is a dereliction
% morphism $d_X$ (the co-unit of the natural transformation of the
% comonad) and 
\begin{proof}
  Since $\C$ is a semantics of linear logic, $E(X)$ comes with a
  $\otimes$-comonoid structure $(E(X),\aff[X]',\cont[X]')$ where
  $\aff[X]'$ is the weakening morphism and $\cont[X]'$ is the
  contraction morphism. Let $\der[X]'$ be the dereliction morphism for
  $X$ of $\C$. Using Theorem~\ref{theo:free}, there exists a morphism
  $\der[X,\ast]'$ of $E(X)\loli !X$. Using Lemma~\ref{lem:Iast} and
  due to the fact that $\C$ is multiset based we obtain that
  $\der[X,\ast]'$ is equal to $\{(x,x)\mid x\in |E(X)|\}$ (the
  inclusion morphism of $E(X)$ in $!X$). Finally, using
  Equation~(\ref{loli:1}), it yields Equation~(\ref{cormax:1}) and
  Equation~(\ref{cormax:2}).
\end{proof}

We shall say that a multiset based semantics of LL in a sub-category
$\C$ of \NCOHK is \emph{non uniform} when the web of the ``of course''
$E$ is the whole set of finite multisets (\ie $|E(X)|=\Mfin(|X|),
\forall X\in \C$).

\begin{corollaire}[sequentiality failure]\label{cor:nodef}
  ~\\
  Each non uniform multiset based semantics of LL in a sub-monoidal
  category $\C$ of \NCOHK fails to reject the 
  morphism
  $\{([\true],\true),([\false],\true),([\true,\false],\true)\}$ of
  type $\bool\rightarrow\bool$. 
\end{corollaire}

%DECOMPRESSE
\begin{proof}
  One easily verifies that $\MultK[\bbbn\backslash\{0,1\}](\{
  [\true],[\false],[\true,\false]\})\cap\Rcohs[!\bool]=\emptyset$ thus
  the morphism above is indeed accepted by the
  $\NCOHK[\bbbn\backslash\{0,1\}]$ semantics.  So this set is \emph{a
    fortiori} a morphism in each \NCOHK semantics and, by maximality of
  the ``of course'', in any multiset based non uniform semantics in a
  sub-monoidal category of \NCOHK.
\end{proof}
%F_DEC
This is a strong negative results since this set cannot be included in
the interpretation of a term of \pcf. Our sentiment is that it will be
the same for any reasonably sequential calculus interpretable in our
semantics of linear logic.  Remark that the very similar morphism
$\{([\true,\true],\true), ([\false,\false],\true),([\true,\false]
,\true)\}$ is the interpretation of $\lambda b.
\Ifthenelse{b}{(\Ifthenelse{b}{\true}{\true})
}{(\Ifthenelse{b}{\true}{\true})}$.

\subsection{Determinism}
\label{sec:determinism}

From now on, we consider that $K$ is a \emph{non-empty} subset of
$\bigK$. In that case the power $\MultK$ is strictly monotone and
preserves disjointness.

\begin{definition}\label{def:nuh}
  Let $\NCohK$ be the full sub-category of $\NCOHK$ whose objects are
  the weakly reflexive \Kcoh spaces.
\end{definition}

Let us recall that being weakly reflexive for a \Kcoh space $X$ means that:
\begin{gather}
  \Rneutre\subseteq \cup_{a\in|X|}\MultK(\{a\}).\label{neutre:1}
\end{gather}

Clearly \NCohK is closed under the orthogonal, additive and
multiplicative constructions. This is also the case for the
exponential construction as easily verified. Indeed assume $X$ is
weakly reflexive and consider a neutral multiset $[x_i\mid i\in I]$ in
$!X$. Then there exists a family $(a^j_i)^{1 \leq j\leq p}_{i\in I}$
such that, for each $i\in I$, $x_i=[a^j_i\mid 1\leq j \leq p]$ and,
for each $1\leq j\leq p$, $[a^j_i\mid i\in I]\in\Rneutre$. So using
weak reflexivity of $X$ we obtain that there exists a family
$(a^j)_{1\leq j\leq p}$ such that $a^j_i=a^j (\forall i, j)$ and
consequently all the $x_i$ are equal.

%So the property
%(\ref{neutre:1}) is preserved by the exponential construction.

% Now
% assume $i$ ranks over $\{1,\ldots,n\}=I$ and consider
% $k\in\bbbn$ such that $ 2\leq k \leq n$. Then using the property
% (\ref{neutre:2}) of $X$ we obtain that the array
% $(a^j_i)_{1\leq i\leq k}^{1\leq j \leq q}$ is as for the
% definition of the neutrality of $(x_i)_{1\leq i\leq k}$ which
% concludes to the preservation of be property (\ref{neutre:2}).

Hence this sub-category is a denotational semantics of propositional
linear logic. Each forgetful functor $U_{K,K'}$ between $\NCOHK$ and
$\NCOHK[K']$ (for $K'\subseteq K$) defines a forgetful functor between
$\NCohK$ and $\NCohK[K']$ having similar properties and for which we
use the same notation $U_{K,K'}$.

% The embedding of the category \Coh (resp. \HC) of usual
% coherent (resp.  hypercoherent) spaces into the category
% $\NCOHK[\{2\}]$ (resp.  $\NCOHK[\bbbn\backslash\{0,1\}]$) is in fact into
% $\NCohK[\{2\}]$ (resp.  $\NCohK[\bbbn\backslash\{0,1\}]$).

\begin{proposition}[determinism]
\label{prop:determinism}
  If $X\in\NCohK$ and if $x$ is a clique of $X$ and $y$ is an
  anti-clique of $X$ (that is a clique of $X\orth$) then
  $\card (x\cap y)\leq 1$.
\end{proposition}

This is a direct consequence of
Proposition~\ref{prop:weakreflexdetermin}.

% \begin{proof}
%   Since $ \MultK(x)\subseteq\Rcoh$ and $\Mult(y)\subseteq\Rincoh$,
%   $\MultK(x\cap y)\subseteq\Rneutre$ and we conclude using
%   Property~\ref{neutre:1}.
% \end{proof}

\begin{rem}
  It is worth remarking that in non-uniform \Kcoh semantics we do not
  have the second part of Property~\ref{property:cohdeterminism}: if
  $(a,c)\in f\comp g$ for $f:X\to Y$ and $g: Y\to Z$ then there exists
  a $b$ such that $(a,b)\in f$ and $(b,c)\in g$ but $b$ is not
  necessarily unique. But uniqueness of $b$ holds again if
  $\MultK(\{(a,c)\}) \subseteq \Rneutre[X\to Z]$ and moreover in that
  case $b$ is such that $\MultK(\{b\})\subseteq\Rneutre[Y]$.
\end{rem}

We are not more interested in non weakly reflexive \Kcoh spaces, and
the category \NCOHK. In the sequel, the most general category will be
\NCohK.

\section{Relating uniform and non uniform semantics}
\label{sec:relunifnonunif}
%\subsection{The new categorical semantics}
We now intend to define uniform \Kcoh semantics and to relate them
with non uniform \Kcoh semantics.

\subsection{The neutral web}
\label{sec:neutralweb}

Proposition~\ref{prop:determinism} can be made more precise since only
certain points can be at the intersection of a clique and an
anti-clique.  These points constitute the \emph{neutral web}.

\begin{definition}
  Let $X\in\NCohK$. We call \bindex{neutral web} of $X$ and we denote
  by $\NK{X}$ (or simply by $\N{X}$) the set $\{a\in |X|\mid
  \MultK(\{a\})\subseteq\Rneutre\}$.
\end{definition}

\begin{exemple}
  For the \Kcoh space $G$ of Example~\ref{ex:G}
  page~\pageref{ex:G} we have: $[a,b]\in\NK{!G}$, if
  $K\subseteq\{2\}$, $[a,b,c]\in\NK{!G}$ and elsewhere
  $[a,b,c]\notin\NK{!G}$.
\end{exemple}

A key result about the neutral web is its behaviour when an ``of
course'' construction is performed:

\begin{lemme}[key lemma]\label{lem:key}
  For $X\in\NCohK$ one has
\[
 \NK{!X}=\{x\in\Mfin(\NK{X})\mid \supp{x}\in\Cl(X)\}
\]
\end{lemme}
\begin{proof}  
  Let $x\in\NK{!X}$. Then for all $k\in K$, there exists a family
  $(a^j_i)^{j\in J}_{1\leq i\leq k}$ such that $[a^j_i\mid j\in J]=x$
  and $[a^j_i\mid 1\leq i\leq k]\in \Rneutre[X]$.  Due to
  Equation~(\ref{neutre:1}), for each $j\in J$, $a^j_1=\ldots =
  a^j_k$. Hence for all $k\in K$, for all $a\in x$,
  $k.[a]\in\Rneutre$. So $\supp{x}\subseteq \NK{X}$.  Each
  $y\in\MultK(\supp{x})$ is a section of the multiset $(\card
  y).[x]\in \Rneutre[!X]\subseteq\Rcoh[!X]$, hence $\supp{x}$ is a
  clique. Thus the left to right inclusion is proved.  Conversely, let
  $x\in\Mfin(\NK{X})$. If $\supp{x}$ is a clique then
  $k.[x]\in\Rcoh[!X]$ for any $k\in K$.  Moreover each of the element
  $a$ of $x$ satisfies $k.[a]\in\Rneutre[X]$ thus $k.[x]\in\Rcoh[!X]$
  for any $k\in K$.  And this proves the right to left inclusion.
\end{proof}

\begin{exemple}
  In $(!G)\orth$, the set $x=\{[a,b],[a,c]\}\subseteq\NK{(!G)\orth}$
  is not a clique if $2\in K$ but is a clique if $2\notin K$. Hence
  $[[a,b],[a,c]]\notin\NK[\{2\}]{!(!G)\orth}$ and
  $[[a,b],[a,c]]\in\NK[\{3\}]{!(!G)\orth}$.
\end{exemple}

The property stated in this lemma teach us that a restriction to the
reflexive subspaces of \Kcoh spaces has good chances to provide us
with a new version of the semantics comparable to the multiset based
coherence space semantics, Equation~\eqref{eq:webofcoursecoh}, when
$K=\{2\}$. This will be successfully shown, among others things, in
the next section. A more direct consequence,
Proposition~\ref{prop:fNK}, is that such a restriction can be
performed at any inductive step of the interpretation. Provided it is
performed at the last step, the resulting reflexive object (in the
case of the interpretation of a formula) or morphism (in the case of
the interpretation of a proof) will be the same.

\begin{definition}
\label{def:N}
  If $X\in\NCohK$, the \bindex{neutral restriction} of $X$ is the
  \emph{sub-space} of $X$ of web $\N{X}$, that is $(\N{X},\Rneutre\cap
  M, \Rcohs\cap M, \Rincohs\cap M)$ where $M=\MultK(\N{X})$, and the
  neutral restriction of a clique $x$ of $X$ is $x\cap\N{X}$. The
  functor $\fNK:\NCohK\to\NCohK$, sometimes simply denoted by $\fN$,
  associates to objects and morphisms their neutral restrictions.
\end{definition}

One easily verifies that $\fNK$ is indeed a functor.

\begin{rem}
  The \Kcoh space $\N{X}$ is reflexive. Moreover it is the maximal
  reflexive subspace of $X$.
\end{rem}

%OPT------
% For each $X\in\NCohK$ the identity morphism $\id[\fNK X]\in\CohK(\fNK
% X, \fNK X)$ can be considered as a morphism of $\NCohK(\fNK X,X)$,
% notation $\id[\fNK X]^{\subseteq}$, or as a morphism of $\NCohK(X,\fNK
% X)$, notation $\id[\fNK X]^{\supseteq }$. For each $f\in\NCohK(X,Y)$,
% $\fNK f$ is then equal to $\id[\fNK X]^{\subseteq}\comp f\comp
% \id[\fNK X]^{\supseteq}$. 

\begin{proposition}\label{prop:fNK}
  The functor $\fNK$ commutes with all the multiplicative additive
  constructions. Moreover $\fNK !=\fNK!\fNK$.
\end{proposition}

\begin{proof}
  The first statement is an obvious consequence of the corresponding
  definitions.
  
  On objects, $\fNK ! =\fNK ! \fNK$ is a consequence of
  Lemma~\ref{lem:key}. Indeed, in the right part of the equality
  stated in this lemma, $\Cl(X)$ can be replaced with $\Cl(\fNK X)$
  since $\supp{x}\subseteq \NK{X}$. This gives $\NK{!X}=\NK{!\fNK X}$
  which is what we wanted. The equality $\fNK != \fNK !\fNK$ on
  morphism is a straightforward consequence of the equality on
  objects.
\end{proof}

\subsection{Uniform $K$-coherence semantics}
\label{sec:uniform-k-coherence}

In this section, we define a uniform $K$-coherence semantics in the
full sub-category of \NCohK which objects are reflexive \Kcoh
spaces. So uniform will be a synonym of reflexive for objects of \NCohK.  

We denote by $\CohK$ the full sub-category of $\NCohK$ whose objects
are the \uKcoh spaces.
% Particular \Kcoh spaces are the ones whose web coincide with their
% neutral web. Theses spaces form a full sub-category of \NCohK.

% \begin{definition}
%   Let $K\subseteq\bbbn\backslash\{0,1\}$. A \bindex{uniform \Kcoh
%     space} is just a \Kcoh space where neutrality coincides with
%   equality: \ie such that $\Rneutre=\cup_{a\in |X|}\MultK(\{a\})$.
%   We denote by $\CohK$ the full sub-category of $\NCohK$ whose objects
%   are the \uKcoh spaces.
% \end{definition}

% Remark that \uKcoh spaces are particular reflexive \Pcoh spaces.
% In a \uKcoh space $X$, any of the relations $\Rcoh$, $\Rcohs$,
% $\Rincoh$, $\Rincohs$ determines the whole structure of the space.

A \uKcoh[\{2\}] space is just an ordinary coherence space.

%$\fNK$ est une projection
The functor $\fNK$ maps $\NCohK$ to $\CohK$ and on $\CohK$, $\fNK$
acts like the identity functor.

Additive and multiplicative constructions of \NCohK preserve uniform
\Kcoh spaces.  This is not the case for the ``of course''
functor. Fortunately, Lemma~\ref{lem:key} gives a clear hint on
what should be the right exponentials for \CohK. 

\begin{definition}\label{def:uexp}
  We define the functor $\ue$ interpreting the ``of course'' in \CohK by
  setting $\ue = \fNK !$.  We denote by $\uwn$ the corresponding ``why
  not'' functor. 
  
  The web of $\ue X$, called the \emph{uniform web}, is then \[|\ue
  X|=\{x\in\Mfin(|X|)\mid \supp{X}\in\Cl(X)\}\] and the coherence of
  $\ue X$ is then given by
\[
M\in\Rcoh[\ue X]\ssi \{m \mid m\msect M\}\subseteq\Rcoh[X].
\]
% $\uwn$ is then equal to $(\ue (\cdot)\orth)\orth$.
\end{definition}

This definition of the exponentials appears as a multiplicities aware
version of the hypercoherences exponentials that have been introduced
in~\cite{hyper}.

As stated by the following theorem, these definitions give rise to a
new class of uniform semantics together with a
straightforward way to extract these interpretations from the
non uniform ones.

\begin{theo}\label{theo:unif}
  For each $K\subseteq\bbbn\backslash\{0,1\}$, \CohK equipped with the
  uniform exponentials and the standard multiplicative additive
  structures of \NCohK is a categorical semantics of linear logic.
  Moreover:
  \begin{enumerate}
  \item the functor $\fNK:\NCohK\to\CohK$ is \emph{logical} which
    means in particular that the \bindex{neutral restriction} of the
    \Kcoh space $\IK{A}$ is the uniform $K$-coherent
    interpretation $\IKu{A}$ of a formula $A$ and that the neutral
    restriction $\IK{\pi}\cap \NK{\vdash\Gamma}$ of the \Kcoh
    interpretation of a proof $\pi$ of a sequent $\vdash\Gamma$ is the
    uniform \Kcoh interpretation $\IKu{\pi}$ of $\pi$;
  \item when $K=\{2\}$ this semantics is exactly the usual
  multiset based coherence semantics.
  \end{enumerate}   
\end{theo}

\begin{proof}
  The multiplicative-additive part of the verification of the fact
  that \CohK is  a semantics of linear logic is easy and relies
  essentially on the fact that $\fN$ commutes to all the additive
  and multiplicative constructions.
  
  The exponential part is not very complicated either. By setting
  $\der[u,X]=\fN(\der[X])$ and $\dig[u,X]=\fN(\dig[X])$ for each
  $X\in\CohK$, we obtain two natural transformations $\der[u]:\fN
  !\dot{\to}\fN \id$ and $\dig[u]:\fN!\dot{\to}\fN!!$ in \CohK.
  
  But $\fN$ is the identity functor on \CohK, $\fN ! = \ue$ and using
  Proposition~\ref{prop:fNK} ($\fN ! =\fN !\fN$ in \NCohK) we obtain
  $\fN !! = \ue\ue$, and also $\fN !!! = \ue\ue\ue$. So $\der[u]$ and
  $\dig[u]$ are in fact natural transformations
  $\der[u]:\ue\dot{\to}\id$ and $\dig[u]:\ue\dot{\to}\ue\ue$.
  
  These two natural transformations endow $\ue$ with a comonad
  structure. In fact we deduce the commutation of the required
  diagrams from the commutation of the corresponding diagrams already
  holding for the comonad $(!,\der,\dig)$ by use of the functor $\fN$.
  The only non-obvious step is then to prove that for each
  $X\in\CohK$, \[\fN\dig[!X]=\dig[u,\ue X]\text{ and }\fN\der[!X]=\der[u,\ue
  X].\]
  This can be done as follows. For all $f\in\NCohK(\fN ! X, !X)$
  one has \[{!f\comp \dig[!X]}=\dig[\fN !  X]\comp f\] hence \[\fN(!f\comp
  \dig[!X])=\fN(\dig[!\fN ! X]\comp f)\] and so 
  \begin{equation}
\fN ! f\comp \fN\dig[!X]=\fN(\dig[\fN ! X])\comp \fN f.\label{eq:1}
\end{equation}
The set $\id[\fN !X]$ is clearly a clique of $\fN !X\loli !X$ and so
it can be seen as a (inclusion) morphism $i$ from $\fN !X$ to $!X$.
We then have the set equalities 
\[\fN !  i=\id[\ue\ue X]\text{ and
}\fN i=\id[\ue X]\]
 so finally by taking $f=i$ in Equation~(\ref{eq:1})
% \[\fN ! f\comp
%  \fN\dig[!X]=\fN(\dig[\fN ! X])\comp \fN f\]
  we obtain the set equality \[\fN\dig[!X]=\fN\dig[\fN!X]\text{ that is }
  \fN\dig[!X]=\dig[u,\ue X].\]  Starting from the equation
  \[\fN(\der[!X]\comp i)=\fN(!i\comp\der[!X])\] one proves
  \[\fN\der[!X]=\der[u,\ue X]\] in the same way.
  
  Using Proposition~\ref{prop:fNK} we obtain the isomorphisms $\ue
  A\otimes\ue B\cong\ue(A\& B)$ and $\ue\top\cong 1$.
  
  $\CohK$ has been proved to be a categorical semantics of linear logic
  and there is nothing more to say for stating that $\fN$ is
  logical.
  
  The comonoid structure of the exponential $\ue$ is then the image of
  the comonoid structure of the exponential $!$ of \NCohK through the
  functor $\fN$. The fact that $(\ue,\der[u])$ is co-free relies
  essentially on the set equality $\der[u,X]=\der[X]$ ($\forall
  X\in\CohK$) which is just a consequence of the fact that all
  singletons are cliques in \CohK. In fact, given a commutative
  comonoid $(A,u_A,\mu_A)$ of \CohK and $f\in\CohK(A,X)$ one has \[\fN
  (f_{\ast})\comp\der[u,X]=\fN (f_{\ast}\comp\der[X])\] for each
  $f\in\CohK(A, X)$ where $f_{\ast}$ is the unique comonoid morphism
  $A\to !X$ such that $f_{\ast}\comp\der[X]=f$. But
  $\fN(f_{\ast}):A\to \ue X$ is also a comonoid morphism. Remark that
  the inclusion morphism $i:\ue X\to !X$ is a comonoid morphism hence
  $\fN(f_{\ast})\comp i:A\to !X$ is a comonoid morphism. We also have
  the set equalities \[\fN(f_{\ast})\comp i=\fN(f_{\ast})\] and, due
  to $\der[X]=\der[u,X]$, \[\fN(f_{\ast})\comp
  i\comp\der[X]=\fN(f_{\ast})\comp\der[u,X]=f.\] By uniqueness of
  $f_{\ast}$, $\fN(f_{\ast})\comp i$ equals $f_{\ast}$, so we finally
  obtain the set equality $f_{\ast}=\fN(f_{\ast})$, and the
  co-freeness of $\ue$ follows.
  
  Finally, $[x,y]\in\Rcoh[\ue X]$ iff $\forall a\in x, \forall b\in y,
  [a,b]\in\Rcoh$ that is, in \CohK[\{2\}], iff $\supp{x+y}$ is a
  clique. So in \CohK[\{2\}] which is the category of coherence
  spaces, $\ue$ is the well-known multiset based exponential of
  coherence spaces.
\end{proof}

Spelling out the categorical definition of the semantics,
the interpretation of linear logic in \CohK is now defined as its
interpretation in $\Rel$ for the multiplicative-additive and identity
groups and with an exponential group similarly defined but using
uniform exponentials and the restriction they induce on the
interpretation of proofs. 

The promotion and the contraction rules cases are subject to the
standard restrictions: in the case of the contraction take only the
$(\gamma,\mu_1+\mu_2)$ such that $\supp{\mu_1+\mu_2}$ is a clique of
$\IK{A}\orth$ and for the promotion, in $f^\dag$, take only the points
such that, for each $i\leq k$, $\supp{\sum_{j\in J} \mu_i^j}\in
\Cl(\IK{A_i}\orth)$. As for usual coherence spaces and
hypercoherences, this condition is sufficient to ensure that
$[a_1,\ldots,a_k]\in \Cl(\IK{A})$ (under the assumption that $f$ is
truly a clique).

\subsection{Multicoherences}
\label{sec:multicoh}

We call the categorical semantics based on $\CohK[\bigK]$ the
\emph{multicoherence semantics}\footnote{General graph theory misses a
  term for such graphs and, contrarily to the \emph{hyper-} situation
  where hypercoherences and hypergraphs are the same,
  \emph{multigraphs} already exist but are not multicoherences.}, we
call \emph{multicoherences} its objects, and we also call
\emph{non uniform multicoherences} the objects of \NCohK[\bigK]. The
only difference between hypercoherences and multicoherences is that
multicoherences take into account the multiplicity of points for the
coherence relation.

\begin{proposition}[sequentiality]\label{prop:defi}
  In the multicoherence semantics, every finite clique of function type
  $!(\bool\with\ldots\with\bool)\loli\bool$
  %in the hierarchy of simple types 
  is sub-definable in \pcf.
\end{proposition}

\begin{proof}
  The proof follows the same scheme as for the usual hypercoherence
  semantics.
\end{proof}

\begin{rem}
  All cliques in the multicoherence semantics are cliques in the coherence
  semantics (this is a consequence of the foliation property).
\end{rem}

\subsection{Non uniform hypercoherences}
\label{sec:hypercoh}

Hypercoherences can be seen as particular multicoherences: the
multicoherences $X$ such that \[\forall
u\in\Rcoh,\supp[-1]{\supp{u}}\subseteq \Rcoh.\]

If $X$ is a non uniform multicoherence having this property for both
the coherence relation and the incoherence relation we say that $X$ is
a \emph{non uniform hypercoherence}. So a non uniform hypercoherence
is indeed simply a weakly reflexive $\Pfinast$-space.
%  such that
% \begin{gather}
%   \Rneutre \subseteq \cup_{a\in |X|} \Pfinast(\{a\})
% \label{eq:weakreflexhyper}
% \end{gather}
But it is more convenient here to present non uniform hypercoherences
as particular non uniform multicoherences.

If $X$ is a non uniform multicoherence, $S(X)$ is the non uniform
hypercoherence defined by
\begin{gather*}
  \Rcoh[S(X)]=\{u\in\MultK[\bigK](|X|)\mid \supp[-1]{\supp{u}}\subseteq\Rcoh\}\\
  \Rneutre[S(X)]=\{u\in\MultK[\bigK](|X|)\mid \supp[-1]{\supp{u}}\subseteq\Rneutre\}
%\text{so }  \Rincohs[S(X)]=\{u\in\Mfin(X)\mid
%  \supp[-1]{\supp{u}}\cap\Rincohs\neq\emptyset\}
\end{gather*}
Remark that the operation $S\ue$ which maps $X$ to $S(\ue X)$ is the
hypercoherence multiset based exponential construction on objects.

\begin{theo}~
\begin{enumerate}
\item The sub-category \NHC of \NCohK[\bigK] of objects the
  non uniform hypercoherences, equipped with the exponential $S!$ on
  objects and acting like $!$ on morphisms is a semantics of linear logic.
  
\item The functor $\fN$ from $\NHC$ to $\HC$, the category of
  hypercoherences, is logical (for the multiset based hypercoherence
  semantics).
  
\item The exponentials $S!$ and $S\ue$ are respectively co-free in
  \NHC and \HC.
\end{enumerate}
\end{theo}

\begin{proof}
  The proof of these statements follows from the proofs of
  Proposition~\ref{prop:ll}, Theorem~\ref{theo:free} and
  Theorem~\ref{theo:unif}.  Just remark that some results can be
  re-used since for each non uniform multicoherence $X$, one has
  $S(!X)=S(!S(X))$ and $\fN(S(X))=S(\fN(X))$.
\end{proof}

Remark that Corollary~\ref{cor:max} applies to $S!$ and $S\ue$.

\begin{exemple}
  The \Kcoh space $G$ of the Example~\ref{ex:G} page~\pageref{ex:G} is
  uniform. And when $K=\bigK$, $G$ is an hypercoherence.  The
  multisets $[a,b]$ and $[a,c]$ are elements of $\N{!G}$. The set
  $x=\{[a,b],[a,c]\}\subseteq \N{!G}$ is a anti-clique of $S(\ue G)$.
  But this set is not an anti-clique (nor a clique) of $\ue G$.  Hence
  each finite multiset of support $x$ is an element of $\N{?S(\ue G)}$
  but not an element of $\N{?\ue G}=\N{?! G}$.
\end{exemple}

For sake of direct usability, we spell out the definition of the non
uniform exponential of hypercoherences on objects directly in the
$\Pfinast$-space setting. The definition of this exponential, denoted
$\nuhce$ is as follows.

If $X$ is a weakly reflexive $\Pfinast$-space then $\nuhce X$ is the
$\Pfinast$-space of web $|\nuhce X| = \Mfin(|X|)$ and such that for
each $x\subfinast |\nuhce X|$
\begin{align}
  \Rincohs[\nuhce X] &\ssi \exists s\in \Rincohs, s \sect x\\
  x\in\Rneutre[\nuhce X] &\ssi \exists \mu, x = \{\mu\} \et \forall
  a\in\mu, \{a\}\in \Rneutre
\end{align}
Of course, $\nuhce X$ is weakly reflexive.

\subsection{Extensional collapses}
\label{sec:collapses}

Consider the situation where a same symmetric monoidal closed category
has two different exponentials defining two different semantics of linear
logic. 

P.-A.~Melli\`es has shown that if there is a \emph{coercion} between
the two exponentials which preserves some structure then the two
semantics will have the same extensional collapse
(\cite{melliescompcollapse}). He uses this result to prove that the
extensional collapse of the multiset-based hypercoherence semantics is
the set-based hypercoherence semantics and he also reproved the same
thing for the coherence spaces semantics (this was already proved by
Barreiro and Ehrhard in~\cite{NunoCollapse}).

This result easily applies to our situation. We then obtain that the
multiset based coherence semantics and the non uniform coherence space
semantics have the same extensional collapse which is the set based
coherence space semantics; the same for hypercoherences; and the same for
multicoherences which we equip with the set based exponential $\se$
defined by:
\begin{gather*}
  |\se X|=\{x\in\Pfin(|X|)\mid x\in\Cl(X)\} \\
  M\in\Rcoh[\se X]\ssi \{m \mid m\msect M\}\subseteq\Rcoh.
\end{gather*}
Of course this last exponential also provides a semantics of linear logic.

We can characterize more precisly the relation between extensional
collapses of uniform and non uniform semantics.

Let $\cal M$ and $\cal M'$ be respectively the non uniform and the
uniform semantics either of coherence spaces, hypercoherence or
multicoherence semantics. Let $\approx$ and $\sim$ be the extensional
PERs respectively on $\cal M$ and $\cal M'$. In what follows $N$
is the neutral restriction functor.

\begin{lemme}
  Let $\sigma$ and $\tau$ be simple types. If $f$ is a clique of
  ${\cal M}(\sigma\to \tau)$ and $x$ is a clique of ${\cal M}(\sigma)$
  then the clique $N(f(x))$ of $\cal M'(\tau)$ is equal to
  $N(f)(N(x))$ and also to $N(f)(x)$.
\end{lemme}

\begin{proof}
  The equality $N(f)(N(x))=N(f)(x)$ is trivial. We only prove
  $N(f(x))=N(f)(N(x))$. Since $Nf\subset f$ and $Nx\subset x$,
  $N(f)(N(x))\subset f(x)$ and since $N$ is monotone
  $N(f)(N(x))=N(N(f)(N(x)))\subset N(f(x))$. Conversely let $b\in
  N(f(x))$ then there exists a $\mu\in\Mfin(x)$ such that $(\mu,b)\in
  f$ and $k[b]$ is neutral for all $k\in K$ ($K=\{2\}$ or $K=\bigK$).
  Since $k[b]$ is neutral and $f$ is a clique $k[\mu]$ is incoherent
  in $!{\cal M}(\sigma)$ but since $x$ is a clique $k[\mu]$ is also
  coherent in $!{\cal M}(\sigma)$. Thus $\mu\in N!{\cal M}(\sigma)$
  and so $(\mu,b)\in Nf$, and $\mu\in\Mfin(Nx)$. This concludes by
  stating $b\in N(f)(N(x))$.
\end{proof}

\begin{lemme}
  If $\sigma$ is a simple type and if $f$ and $g$ are cliques of
  $\cal M(\sigma)$ then \[f \approx_{\sigma} g \ssi  Nf \sim_{\sigma} Ng.\]
\end{lemme}

\begin{proof}
  This is trivially true on basis type (on basis type $N$ acts as the
  identity functor). Suppose this is true for types $\sigma$ and
  $\tau$. We prove the property for the type $\sigma\to\tau$. Let $f
  \approx_{\sigma\to \tau} g$ and let $x \sim_{\sigma} y$. Since
  $Nx=x$ and $Ny=y$, by induction hypothesis, $x \approx_{\sigma} y$.
  Hence $f(x)\approx_{\tau} g(y)$ and so, by induction hypothesis,
  $(Nf)(x)\sim_{\tau} (Ng)(y)$. So $f \approx_{\sigma\to \tau} g$
  implies $Nf \sim_{\sigma\to \tau} Ng$. Conversely let $f$ and $g$ be
  two cliques of ${\cal M}(\sigma\to\tau)$ such that $Nf
  \sim_{\sigma\to\tau} Ng$. Let $x \approx_{\sigma} y$. Then, by
  induction hypothesis, $Nx \sim_{\sigma} Ny$.  Hence $Nf(Nx)
  \sim_{\tau} Ng(Ny)$ and $N(f(x)) \sim_{\tau} N(g(y))$. And, by
  induction hypothesis, $f(x) \approx_{\tau} g(y)$. This concludes
  proving the lemma by stating that $Nf \sim_{\sigma\to \tau} Ng$
  implies $f \approx_{\sigma\to \tau} g$.
\end{proof}

\begin{theo}
  The neutral functor $N$ defines a one to one correspondence between
  the extensional collapse of $\cal M$ and the extensional collapse of
  $\cal M'$.
\end{theo}

This is a direct consequence of the last lemmas.

Finally using the fact that the set based hypercoherence and
multicoherence semantics are both extensional we show that
hypercoherences and multicoherences are extensionally different by
exhibiting, in Example~\ref{ex:diffextcoll}, a relation at a functional
type which is a clique in one of two semantics but not in the other (this
example was originally designed to exhibit a set which is a clique in
the set-based hypercoherence semantics but not in the coherence
spaces semantics).

\begin{exemple}\label{ex:diffextcoll}
  For the hypercoherence $G$ of our last examples above, one has that
  $\{a,b\}$ and $\{c\}$ are elements of $|\se G|=|S(\se G)|$. Moreover
  $(\{a,b\},\true)$ and $(\{c\},\false)$ are elements of $|\se
  G\loli\bool|=|S(\se G)\loli\bool|$. The relation
  $F=\{(\{a,b\},\true),(\{c\},\false)\}$ is a clique of the
  hypercoherence $S(\se G)\loli \bool$. But $F$ is not a clique of the
  multicoherence $\se G\loli \bool$, since $[\{a,b\},\{c\}]$ is
  coherent but $[\true, \false]$ is strictly incoherent.
\end{exemple}

%   The hypercoherence $G$ of our examples is indeed a
%   subspace of the set based hypercoherence interpretation of
%   $(\bool\times\bool\times\bool\to\bool)\to\bool$ which is equal to
%   its set based multicoherence interpretation. 
\section{Conclusion and further works}

\subsection{Static interactivity}
\label{sec:staticinteractivity}
The present work gives some strong evidence that static uniformity is
a matter of restriction to possible results of computation, through
interactions and especially through interactions in a closed case:
between a \emph{proof} of $A$ and a \emph{proof} of $A\orth$. We
further argue on this point by adopting basic ideas of the interactive
point of view on computation developped in Girard's
ludics~\cite{locus}.

The main idea is to consider the linear logic system extended with new
rules (such that the system still enjoys cut-elimination). Then a
formula $A$ and its linear negation $A\orth$ can be both
provable. Hence we can provoke interactions through cut elimination
between proofs of $A$ and proofs of $A\orth$. If we further require
that the extended logical system admits one of the deterministic non
uniform semantics we presented then the consequence of determinism is
that such an interaction involves at most one point.

Let suppose that we add two para-rules to linear logic:
\[
\GIVEUPR
\DP\qquad
\DIVERGER{\Gamma}
\DP
\]
a \emph{give up} (this \emph{para-}rule roughly corresponds to the
\emph{daemon} of Girard's Ludics~\cite{locus}) rule and a
\emph{divergence} rule whose respective interpretations in the
relational semantics are a singleton (the unique point in the unit
context $\bot$) and the empty set. It is easy to extend the cuts
elimination procedure for this two rules and to check that for
instance the relational semantics extends into a semantics of the
strongly normalizing calculus we then obtain. In this setting a
formula $A$ and its dual $A\orth$ are always both provable.

When we apply a cut rule between a proof $\pi'$ of $A$ and a proof
$\pi''$ of $A\orth$, the proof $\pi$ we obtain normalizes into a proof
of the empty sequent. And there is only two cut free proofs of the
empty sequent: one is an instance of the \emph{give up} rule and one
is an instance of the \emph{divergence} rule, with an empty
context. One easily verifies that the resulting cut free proof is a
\emph{give up} (resp. a \emph{divergence}) iff the relational
interpretation of $\pi'$ and $\pi''$ have a non empty intersection
(resp. empty). If the interpretation is not empty we shall say that
$\pi'$ and $\pi''$ interact. From the point of view of the bipartite
relational semantics the \emph{give up} rule is not valid (since it
introduces a sequent interpreted by a negative point) and there can be
no interaction between a proof of $A$ and a proof of $A\orth$. We
think that this is the fundamental reason which makes possible a
uniform semantics where so much proofs have empty interpretations:
uniformity is a restriction to possible interactions and if no
interaction is possible uniformity empties things.

Remark that \emph{give up} and \emph{divergence} are valid rules in
the others coherence like semantics we present in this paper. In
particular, the property of determinism of non uniform semantics tell
us that when $\pi'$ and $\pi''$ interact there is only one result of
this interaction (there is only one point in the intersection). This
seems difficult to prove directly in the relational semantics without
introducing non uniform coherence relations.  Moreover the result of
interaction is always in the neutral webs of the various semantics
(coherence spaces, hypercoherences, multicoherences), hence, in the
extended linear logic we use here, a part of the web is never visited
by closed interactions. This will be the case in any extension of
linear logic which admits one of these semantics.

In fact, one can imagine new para-rules such that there can be
interactions on any points of the relational semantics. It has to be
checked if useful, but as Curien suggested us, adding a \emph{sum}
rule like :
\[
\AXC{$\vdash\Gamma$}\AXC{$\vdash\Gamma$}
\RL{(\emph{sum})}\BIC{$\vdash\Gamma$}
\DP
\]
interpreted by a union in the relational semantics, will certainly
gives a semantics of this kind. But determinism will be lost.

\subsection{Extending non uniform static semantics}
Using the co-free exponential for non uniform static semantics have
led to a comfortable situation where non uniform semantics are
deterministic and strongly related with the uniform semantics. A
(still) open question is: can the general construction Bucciarelli and
Ehrhard introduced~\cite{phaseexp} be modified so as to directly
obtain the co-free exponentials in a general way? Another related
issue concerns full completeness for static semantics.  Ehrhard proves
a completeness theorem in an indexed linear logic
framework~\cite{Ehrhard00}. A better understanding of the co-freeness
issue in indexed linear logic may help in connecting his result with
usual static semantics (hypercoherences and coherence spaces).

In an unpublished work, G.~Winskel has introduced a generalization of
hypercoherences. It may be interesting to adapt this generalization to
a non uniform framework including the semantics we present here.

\bibliographystyle{alpha}
\bibliography{biblio}

\begin{thebibliography}{{Bru}01}

\bibitem[AC98]{AmadioCurien}
Roberto~M. Amadio and Pierre-Louis Curien.
\newblock {\em Domains and lambda-calculi}, volume~46 of {\em Cambridge Tracts
  in Theoretical Computer Science}.
\newblock Cambridge University Press, Cambridge, july 1998.

\bibitem[AJM94]{ajm}
Samson Abramsky, Radha Jagadeesan, and Pasquale Malacaria.
\newblock Full abstraction for {PCF}.
\newblock In {\em Theoretical Aspects of Computer Software}, pages 1--15, 1994.

\bibitem[BC82]{cds}
G\'erard Berry and Pierre-Louis Curien.
\newblock Sequential algorithms on concrete data structures.
\newblock {\em Theoretical Computer Science}, 20:265--321, 1982.

\bibitem[BE94]{seqext}
Antonio Bucciarelli and Thomas Ehrhard.
\newblock Sequentiality in an extensional framework.
\newblock {\em Information and Computation}, 110(2), 1994.

\bibitem[BE97]{NunoCollapse}
Nuno Barreiro and Thomas Ehrhard.
\newblock Anatomy of an extensional collapse.
\newblock Available by ftp at \texttt{iml.univ-mrs.fr/pub/ehrhard/}, 1997.

\bibitem[BE00]{phasemall}
Antonio Bucciarelli and Thomas Ehrhard.
\newblock On phase semantics and denotational semantics in
  multiplicative-additive linear logic.
\newblock {\em APAL}, 102(3):247--282, 2000.

\bibitem[BE01]{phaseexp}
Antonio Bucciarelli and Thomas Ehrhard.
\newblock On phase semantics and denotational semantics: the exponentials.
\newblock {\em Annals of Pure and Applied Logic}, 109(3):205--241, 2001.

\bibitem[Bie95]{whatillcatmod}
G.~M. Bierman.
\newblock What is a categorical model of intuitionistic linear logic?
\newblock In M.~Dezani, editor, {\em Proceedings of Conference on Typed lambda
  calculus and Applications}. Springer-Verlag LNCS 902, 1995.

\bibitem[Bou03]{nuhc}
Pierre Boudes.
\newblock Non uniform hypercoherences.
\newblock In Rick Blute and Peter Selinger, editors, {\em Proceedings of CTCS
  2002, Electronic Notes in Theoretical Computer Science}, volume~69. Elsevier,
  2003.

\bibitem[Bou04]{pbg}
Pierre Boudes.
\newblock Projecting games on hypercoherences.
\newblock In {\em ICALP}, volume 3142 of {\em Lecture Notes in Computer
  Science}, pages 257--268. Springer, 2004.

\bibitem[Bou05]{gamesexperiments}
Pierre Boudes.
\newblock Desequentialization of games and experiments in proof-nets.
\newblock pr\'e-publication IML 2005-1, 2005.

\bibitem[{Bru}01]{alexphd}
A.~{Bruasse-Bac}.
\newblock {\em Logique lin{\'e}aire index{\'e}e du second ordre}.
\newblock PhD thesis, Universit{\'e} Aix-Marseille II - M{\'e}diterran{\'e}e,
  2001.

\bibitem[CCF94]{CartwrightCurienFelleisen94}
Robert Cartwright, Pierre-Louis Curien, and Matthias Felleisen.
\newblock Fully abstract semantics for observably sequential languages.
\newblock {\em Information and Computation}, 111(2):297--401, 1994.

\bibitem[Ehr93]{hyper}
Thomas Ehrhard.
\newblock Hypercoherences: a strongly stable model of linear logic.
\newblock {\em Mathematical Structures in Computer Science}, 3, 1993.

\bibitem[Ehr99]{relpcf}
Thomas Ehrhard.
\newblock A relative definability result for strongly stable functions and some
  corollaries.
\newblock {\em Information and Computation}, 152, 1999.

\bibitem[Ehr00]{sp}
Thomas Ehrhard.
\newblock Parallel and serial hypercoherences.
\newblock {\em Theoretical computer science}, 247:39--81, 2000.

\bibitem[Ehr03]{Ehrhard00}
Thomas Ehrhard.
\newblock A completeness theorem for symmetric product phase spaces.
\newblock To appear in The Journal of Symbolic Logic, 2003.

\bibitem[Ehr05]{finiteness}
Thomas Ehrhard.
\newblock Finiteness spaces.
\newblock {\em Mathematical Structures in Computer Science}, 15(4):615--646,
  2005.

\bibitem[Gir87]{linearlogic}
Jean-Yves Girard.
\newblock Linear logic.
\newblock {\em Theoretical Computer Science}, 50:1--102, 1987.

\bibitem[Gir96]{ondenot}
Jean-Yves Girard.
\newblock On denotational completeness.
\newblock manuscript, 1996.

\bibitem[Gir01]{locus}
Jean-Yves Girard.
\newblock Locus {S}olum.
\newblock {\em Mathematical Structures in Computer Science}, 11(3):301--506,
  June 2001.

\bibitem[Lon02]{Longley}
{J.R.} Longley.
\newblock The sequentially realizable functionals.
\newblock {\em Annals of Pure and Applied Logic}, 117(1-3):1--93, 2002.

\bibitem[Mel04]{melliescompcollapse}
Paul-Andr\'e Melli\`es.
\newblock Comparing hierarchies of types in models of linear logic.
\newblock {\em Information and Computation}, 189(2):202--234, March 2004.

\bibitem[Mel05]{melliesseqext}
Paul-Andr{\'e} Melli{\`e}s.
\newblock Sequential algorithms and strongly stable functions.
\newblock {\em Theoretical Computer Science}, 343(1-2):237--281, 2005.

\bibitem[Pag06]{pagani06mscs}
Michele Pagani.
\newblock Proofs, denotational semantics and observational equivalences in
  multiplicative linear logic.
\newblock {\em Mathematical Structures in Computer Science}, ??(??):??, 2006.
\newblock To appear.

\bibitem[{Tor}00]{phd:Lorenzo}
Lorenzo {Tortora de Falco}.
\newblock {\em R\'eseaux, coh\'erence et exp\'eriences obsessionnelles}.
\newblock Th\`ese de doctorat, Universit\'e Paris VII, 2000.

\bibitem[vO97]{VanOosten}
Jaap van Oosten.
\newblock A combinatory algebra for sequential functionals of finite type.
\newblock Technical Report 996, University of Utrecht, 1997.

\end{thebibliography}

\end{document}